\documentclass[11pt]{article}
\usepackage{amssymb}
\usepackage{epsfig}
\usepackage{psfrag}
\def\nn{\nonumber}

\def\la{\langle}
\def\ra{\rangle}
\def\eps{\varepsilon}
\def\l{\left}
\def\r{\right}

\def\beq{\begin{equation}}
\def\eeq{\end{equation}}
\def\bea{\begin{eqnarray}}
\def\eea{\end{eqnarray}}
\def\barr{\begin{array}}
\def\earr{\end{array}}
\def\be{\begin{equation}}
\def\ee{\end{equation}}
\def\bc{\begin{center}}
\def\ec{\end{center}}

\def\ov{\overline}
\def\cL{{\cal L}}
\def\cC{{\cal C}}
\def\cH{{\cal H}}

\def\de{\partial}

\def\op{{\mathcal{O}}}

\def\ChPT{$\chi$PT}
\def\HBChPT{HB$\chi$PT}
\def\vff{vff}
\begin{document}
\begin{titlepage}
\begin{flushright}
HUTP-06/A0010
\end{flushright}
\vskip 2cm
\begin{center}
{\LARGE \bf Chiral Corrections to the Hyperon Vector Form Factors} 
\vskip1cm {\Large  Giovanni Villadoro}\\ \vspace{.5cm}
{\normalsize {\large \sl
Jefferson Physical Laboratory, Harvard University, \\ [3pt]
Cambridge, Massachusetts 02138, USA}} \\ [24pt]
{\sl E-mail: villador@physics.harvard.edu}

\vskip2.0cm {\bf Abstract\\[10pt]} \parbox[t]{\textwidth}{{
We present the complete calculation of the SU(3)-breaking corrections 
to the hyperon vector form factors up to $\op(p^4)$ in the
Heavy Baryon Chiral Perturbation Theory. Because of the Ademollo--Gatto theorem,
at this order the results do not depend on unknown low energy constants 
and allow to test the convergence of the chiral expansion. 
We complete and correct previous calculations and find
that $\op(p^3)$ and $\op(1/M_0)$ corrections are important.
We also study the inclusion of the decuplet degrees of freedom,
showing that in this case the perturbative expansion is jeopardized.
These results raise doubts on the reliability of 
the chiral expansion for hyperons.}}
\end{center}
\vspace{4cm}

%
\end{titlepage}
\baselineskip 5 mm
\section{Introduction}
At present, the most precise determination of the Cabibbo-Kobayashi-Maskawa (CKM) 
matrix element $V_{us}$ \cite{CKM} is obtained
from kaon semileptonic decays (for recent reviews see e.g. \cite{Vus} and references therein). 
In this case, the uncertainties from both experiments and theory
allow to reach the percent precision, which could be further improved in the near future.
The experiments allow to extract with good accuracy the product 
$|V_{us}\cdot f_+(0)|^2$ where $f_+(0)$ is the vector form factor (\vff) at zero momentum transferred.
Because of the conservation of the vector current, the latter is known up to SU(3)-breaking corrections, 
which are suppressed because of the Ademollo-Gatto theorem~\cite{AG}. 
The leading contribution to these corrections can be determined univocally in 
Chiral Perturbation Theory~\cite{glff,lr}, while the subleading corrections can been 
estimated via quark models \cite{lr}, large-$N_c$ \cite{cirNc} 
or in a model independent way via Lattice QCD
simulations \cite{kl3,kl3hyplat,unquench}.

Since $V_{us}$ together with $V_{ud}$ allow the most stringent test of CKM unitarity,
it would be interesting to have other independent estimates for these quantities.

It has been recently pointed out \cite{csw} that $V_{us}$ can be extracted also from hyperon semileptonic
decays in a similar way as from kaon decays. In particular, experimental
data can be combined to extract the product $|V_{us}\cdot f_1(0)|^2$ where $f_1(0)$ is the hyperon vector
form factor at zero momentum. As for the kaon case, Ademollo-Gatto theorem protects $f_1(0)$ from
the leading SU(3)-breaking corrections and make these decays, at least in principle, good
alternatives for the extraction of $V_{us}$. 
There exist several estimates for $f_1(0)$ using quark models \cite{qmodel},
large-$N_c$ \cite{largeN} and chiral expansions \cite{Krause,al,Kaiser}. However, they
disagree between each other, and a model independent 
estimate becomes mandatory (see also \cite{mp} for a recent discussion).
In ref.~\cite{kl3hyplat,hyplat}, a lattice study has been
implemented demonstrating that the method of \cite{kl3} for the extraction of SU(3)-breaking 
corrections from the lattice can be performed also in the hyperon case. 
However, because of the limitations of the actual numerical simulations it is necessary
to estimate the leading chiral corrections to drive the extrapolation
to low quark masses. We thus need an estimate for the leading chiral corrections to
the hyperon \vff. 
However, existing calculations \cite{Krause,al,Kaiser} in (Heavy) Baryon Chiral Perturbation Theory 
(H)B$\chi$PT \cite{jm1} for these amplitudes are not complete (and disagree between each other). 

In this paper we perform the full $\op(p^4)$ calculation in \HBChPT\ of hyperon decays. 
We find that the subleading $\op(p^4)$ contributions (including $1/M_0$ relativistic corrections)
are important and cannot be neglected. The results present cancellations
between different contributions that make the estimates strongly dependent on the low
energy parameters (masses and couplings).
We also study the inclusion of decuplet degrees of freedom in the \HBChPT\ formalism
developed in \cite{jm2,hhk}, finding huge contributions that make the calculation unreliable.
Because of these results, \HBChPT\ does not seem to be of help for the determination
of $f_1(0)$ and lattice simulations closer to the chiral limit have to be performed to
obtain a model independent estimate of these quantities.

The paper is organized as follows. In section~\ref{sec:eft} we review 
\HBChPT\ and fix our notations. In section~\ref{sec:vff} we present the
results for the \vff, in particular 
$\op(p^2)$, $\op(1/M_0)$ and $\op(p^3)$ corrections are studied
in subsections~\ref{sec:op2}, \ref{sec:opM} and \ref{sec:op3oct} 
respectively and the final discussion on octet contributions is given 
in subsection~\ref{sec:finoct}.  In section~\ref{sec:dec} we present
our results for the decuplet contributions.
The conclusions are given in 
section~\ref{sec:concl}. Finally, in the appendices~\ref{sec:app1} and \ref{sec:app2}
we give the explicit expressions for the $\op(p^3)$ octet and decuplet
corrections respectively.

\section{The effective field theory for (heavy) baryons}
\label{sec:eft}
Chiral Perturbation Theory (\ChPT) is the effective field theory 
for the lightest mesons of QCD. Since these states
are the quasi-Goldstone modes of the spontaneously-broken
SU(3)$_L\times$SU(3)$_R$ chiral symmetry, 
a perturbative expansion can be performed
as long as energies below the QCD scale ($\sim m_\rho$) are considered \cite{weinberg}.
The symmetries of the underlying theory (QCD) fix completely the form of the 
interactions \cite{ccwz}. At each order of the expansion,
the high energy dynamics is encoded in a finite number of effective couplings.
These low energy constants (LECs) can be estimated using
experiments or with non-perturbative approaches, such as Lattice QCD. 


In principle, \ChPT\ can be extended to include fermionic degrees of freedom,
(e.g. baryons), whose properties under chiral transformations fix their 
couplings to mesons. The lightest baryons, however, lie above the regime
of validity of \ChPT\ and a naive inclusion of these states would break the power counting.
A way to overcome this problem has been found in \cite{jm1} by adapting the method 
for heavy quarks of ref.~\cite{georgi}. The Heavy 
Baryon Chiral Perturbation Theory (\HBChPT) of ref.~\cite{jm1} is obtained by integrating out 
the heavy components of the fermionic fields that ruin the power counting, i.e.
considering baryons in the non-relativistic limit. In this way, only light degrees of freedom are dynamical
and the power counting is consistent at all orders of the chiral expansion. In \HBChPT\ we thus have
a double expansion: the chiral expansion in powers of $p/\Lambda_\chi$ ($\Lambda_\chi\sim1$~GeV)
and the heavy baryon expansion in powers of $p/M_B$. Since $M_B\sim\Lambda_\chi$ the 
two expansions can be joined together. As we will see, however, the degree of convergence of the perturbative 
series is rather poor and higher order corrections give important contributions. 

In order to fix our convention we will report here the effective Lagrangian 
for baryons and mesons:
\beq
\cL=\cL_{M}^{(2)}+\cL_{B}^{(1)}+\cL_{B}^{(2)}+\cL_{B}^{(3)}+\dots
\eeq

The purely mesonic sector reads \cite{gl}:
\beq \label{eq:L2M}
\cL_M^{(2)}=\frac{f^2}{2} \la A^\mu A_\mu +\rho\, \chi \ra \,,
\eeq
where $\la\dots\ra$ indicates the trace on flavor indices and
\beq
A_\mu=\frac12 i \l ( \xi \de_\mu \xi^\dagger -\xi^\dagger \de_\mu \xi \r) \,, \qquad 
	\chi=\frac12 \l ( \xi^\dagger M \xi^\dagger+ \xi M \xi \r) \,, \qquad
M={\rm diag}\l (m_u,m_d,m_s\r)\,,
\eeq
\beq
\xi=e^{i \Phi/f} \,, \qquad 
\Phi=\l ( \barr{ccc} \frac{\pi^0}{\sqrt2}+\frac{\eta}{\sqrt6} & \pi^+& K^+ \\ 
						\pi^- &  -\frac{\pi^0}{\sqrt2}+\frac{\eta}{\sqrt6}& K^0 \\
						K^- & \ov K^0 & -\frac{2}{\sqrt6}\eta \earr \r)\,.
\eeq
The only parameters appearing in $\cL_{M}^{(2)}$ are the meson decay constant $f$ (normalized so that
$f_\pi \simeq 132$~MeV) and the meson masses, which in the SU(2) limit ($m_\ell=m_u=m_d$) read:
\bea
m_\pi^2&=&2\,\rho\, m_\ell\,, \nn \\
m_K^2&=&\rho\, \l( m_\ell+m_s\r) \,, \nn \\
m_\eta^2&=&\frac43\, m_K^2-\frac13\, m_\pi^2 \,.
\eea
The leading order $\op(p)$ \HBChPT\ Lagrangian is \cite{jm1}:
\beq \label{eq:L1B}
\cL_{B}^{(1)}=\la \bar B \,i\, v^\mu D_\mu B \ra +2 D\, \la \bar B\, S^\mu\, \{A_\mu,B\}\ra
+2 F\,\la \bar B\, S^\mu\, [A_\mu,B]\ra\,,
\eeq
where 
\beq
D_\mu B=\de_\mu B-i [V_\mu,B]\,, \qquad V_\mu = \frac12 i \l ( \xi \de_\mu \xi^\dagger +\xi^\dagger \de_\mu \xi \r)\,,
\eeq
\beq
B=\l ( \barr{ccc} \frac{\Sigma^0}{\sqrt2}+\frac{\Lambda}{\sqrt6} & \Sigma^+& p \\ 
						\Sigma^- &  -\frac{\Sigma^0}{\sqrt2}+\frac{\Lambda}{\sqrt6}& n \\
						\Xi^- & \Xi^0 & -\frac{2}{\sqrt6}\Lambda \earr \r)\,,
\eeq
$v^\mu=p^\mu/M_0$ is the four-velocity of the baryon and $S^\mu=i\gamma_5\sigma^{\mu\nu}v_\nu/2$ is 
the spin operator.
$F$ and $D$ are the axial couplings, whose physical values are known with good accuracy \cite{csw}.
From eq.~(\ref{eq:L1B}) we can read the baryon propagator:
\beq
\frac{i}{k\cdot v+i\,\eps}\,,
\eeq
where $k\sim \op(p)$ is the off-shell momentum carried by the light field component $B$
and, at this order, is related to the momentum $p^\mu$ of the physical baryon 
via the relation 
\beq
p^\mu=M_0\, v^\mu +k^\mu\,,
\eeq 
where $M_0$ is the baryon mass in the chiral limit and $p^2=M_0^2$ on-shell.

At the next order we have $\cL_B^{(2)}$, which receives contributions both from genuine $\op(p^2)$ 
chiral corrections and from $1/M_0$ corrections:
\beq \label{eq:L2B}
\cL_B^{(2)}=\cL_B^{(2)}(p^2)+\cL_B^{(2)}(1/M_0)\,,
\eeq
\bea
\cL_B^{(2)}(p^2)&=&\sigma\, \la \chi \ra\, \la \bar B B \ra
+2\, b_D\, \la \bar B\, \{\chi, B\}\ra+2\, b_F\, \la \bar B\, [\chi, B]\ra + \dots 
	\label{eq:LB2p3} \\
\cL_B^{(2)}(1/M_0)&=&\frac{1}{2M_0}\, \la \bar B \,\l [(v^\mu \de_\mu)^2 - \de^\mu \de_\mu\r]  B \ra \nn \\
 	&& + \frac{i\,v^\nu}{M_0} D\, \la \de_\mu \ov B\, S^\mu\, \{ A_\nu,B\}-\ov B\, S^\mu\, \{ A_\nu, \de_\mu B\} \ra\nn \\
	&& + \frac{i\,v^\nu}{M_0} F\, \la \de_\mu \ov B\, S^\mu\, [ A_\nu,B]-\ov B\, S^\mu\, [ A_\nu, \de_\mu B] \ra + \dots 
	\label{eq:LB2M}
\eea
where we wrote explicitly only those operators giving a non-vanishing contribution to
the hyperon \vff. The three terms in eq.~(\ref{eq:LB2p3}) give the 
lowest order chiral corrections $\delta M_B\sim \op(p^2)$ to the baryon mass $M_0$:
\beq
\frac{i}{k\cdot v-\delta M_B+i\eps}\simeq \frac{i}{k\cdot v+i\eps}+\frac{i}{k\cdot v+i\eps}( - i\delta M_B) \frac{i}{k\cdot v+i\eps}\,,
\eeq
and read:
\bea
\delta M_N&=& 2\, b_F\, (m_s - m_\ell) - 2\, b_D\, (m_\ell + m_s) - \sigma \,(2\, m_\ell  + m_s) \,, \nn \\
\delta M_\Sigma&=& -4\, b_D\, m_\ell - \sigma \,(2\, m_\ell  + m_s) \,, \nn \\
\delta M_\Lambda&=& -\frac{4\, b_D}{3}\, (m_\ell+2\,m_s)- \sigma \,(2\, m_\ell  + m_s) \,, \nn \\
\delta M_\Xi&=&  - 2\, b_F\, (m_s- m_\ell) - 2\, b_D\, (m_\ell  + m_s) - \sigma \,(2\, m_\ell + m_s) \,.
\eea
The terms in the first line of eq.~(\ref{eq:LB2M}) correspond to the $1/M_0$ corrections
to the baryon propagator:
\beq
\frac{i}{/ \hspace{-6pt} p-M_0+i\eps}
\to\frac{i}{k\cdot v+i\eps}+\frac{i}{k\cdot v+i\eps}\l (i\frac{k^2-(k\cdot v)^2}{2 M_0} \r) \frac{i}{k\cdot v+i\eps}\,,
\eeq
while the terms in the second and third line are the relativistic corrections to
the leading order (LO) axial couplings of eq.~(\ref{eq:L1B}).
Note that in $\cL_B^{(2)}(1/M_0)$ no new LECs appear since 
all the coefficients are fixed by Lorentz symmetry. 
On the other hand, the coefficients of the operators in $\cL_B^{(2)}(p^2)$ are undetermined. 
However, since in the calculation of the \vff\ at $\op(p^4)$
only the baryon mass-shift operators of eq.~(\ref{eq:LB2p3}) will contribute, 
no unknown LEC appears at this order.

Since each meson loop corresponds to corrections of $\op(p^2)$, operators 
of order $\op(p^3)$ and $\op(p^4)$ of the \HBChPT \ Lagrangian 
can only enter at tree level in the calculation 
of the \vff\ at $\op(p^4)$. However, because of the
Ademollo-Gatto theorem, the first tree level non-vanishing contribution 
starts at $\op(p^5)$ (since the LO is $\op(p)$), therefore only the operators in 
${\cal L}_B^{(1)}$ and ${\cal L}_B^{(2)}$ 
will contribute. As a by-product, in the \vff\ 
at $\op(p^4)$, loop contributions are finite and no unknown LEC appears.

\section{Vector form factors for hyperons}
\label{sec:vff}
The vector form factor $f_1(q^2)$ for baryons is defined via the matrix element
of the SU(3) vector current as follows:
\beq \label{eq:vecmatrx}
\langle B_2 | V^\mu| B_1\rangle=\overline B_2(p_2) \left [ \gamma^\mu f_1(q^2) 
	- i\frac{\sigma^{\mu\nu}q_\nu}{M_1+M_2} f_2(q^2) +\frac{q^\mu}{M_1+M_2} f_3(q^2)\right ] B_1(p_1) \,.
\eeq
where $q^\mu=p_1^\mu-p_2^\mu$.
In the SU(3) limit the \vff\ at zero momentum $f_1(0)$ 
are fixed by the conservation of the SU(3)$_V$ charge. Out of the SU(3) limit 
Ademollo-Gatto theorem states that linear corrections in the breaking ($m_s-m_{\ell}$) vanish, 
i.e.
\beq
f_1(0)=f_1^{SU(3)}(0)+\op(m_s-m_{\ell})^2\,.
\eeq

The expression for the vector current in \HBChPT\ can be extracted from eqs.~(\ref{eq:L2M}),
(\ref{eq:L1B}) and (\ref{eq:L2B}) by coupling to the system an external vector current 
and by varying the action with respect to it. 
The \vff\ at zero momentum can thus be extracted by looking at
the terms proportional to $v^\mu$ (see \cite{em}) in the heavy baryon limit 
of the matrix element of eq.~(\ref{eq:vecmatrx}).

In the following, we will parametrize chiral corrections to the \vff\ as follows:
\beq \label{eq:not}
f_{1}(0)=\alpha^{(1)}\l [1+ \alpha^{(2)} + \l ( \alpha^{(3)} +\alpha^{(1/M)}\r)+\dots\r]\,.
\eeq
$\alpha^{(1)}=f_1^{SU(3)}(0)$ is the tree-level $\op(p)$ amplitude. $\alpha^{(2)}$ is
the one-loop correction ($\op(p^3)$) and it is $\op(p^2)$ with respect to $\alpha^{(1)}$. 
$\alpha^{(3)}$ and $\alpha^{(1/M)}$ are respectively the $\op(p^4)$ chiral contribution 
and the leading $1/M_0$ corrections, and both are $\op(p^3)$ with respect to $\alpha^{(1)}$.

The tree-level amplitudes give just the SU(3)$_V$ charges and read:
\beq
\alpha^{(1)}_{\Sigma^- n}=-1\,,\qquad
\alpha^{(1)}_{\Lambda p}=-\sqrt{\frac32}\,,\qquad
\alpha^{(1)}_{\Xi^-\Lambda}=\sqrt{\frac32}\,,\qquad
\alpha^{(1)}_{\Xi^-\Sigma^0}=\frac{1}{\sqrt2}\,.
\eeq

\begin{figure}[t]
\bc \epsfig{file=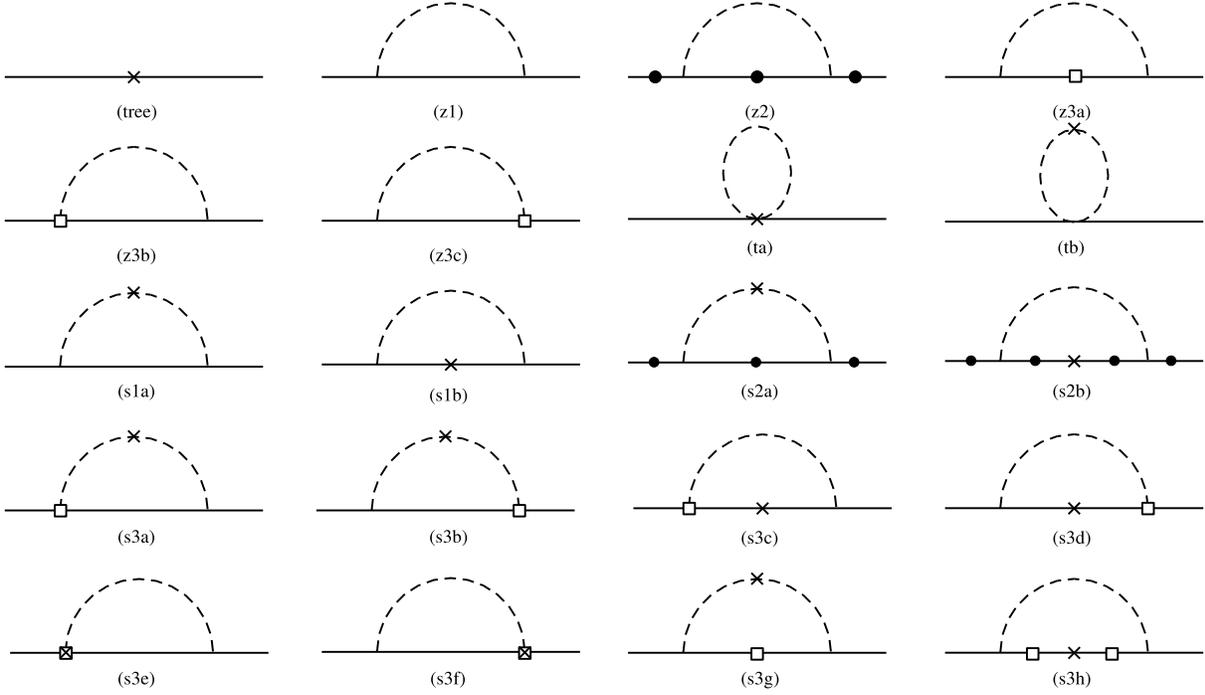,width=\textwidth} \ec 
\caption{\it Feynman diagrams of octet contributions to the \vff\ up to $\op(p^4)$ in the chiral expansion.
Solid (dashed) lines are baryons (mesons), crosses are vector current insertions, filled circles are $\op(p^2)$ operators,
empty boxes are $1/M_0$ corrections and crossed boxes are $1/M_0$ corrections to the vector current. 
The graphs with multiple insertions of $\op(p^2)$ or $1/M_0$ operators are a 
short-hand notation and correspond to the sum of the corresponding graphs 
with single insertions.}
\label{fig:graph8}
\end{figure}


%
\subsection{$\op(p^2)$ corrections}
\label{sec:op2}
The leading SU(3)-breaking corrections occur at one-loop in the chiral expansion.
The relevant diagrams, which give a non-vanishing contribution at this order,
are those labeled as (z1), (ta), (tb), (s1a) and (s1b) in figure~\ref{fig:graph8}. 
These diagrams can be divided into two classes: tadpoles [(ta),(tb)] and sunsets [(z1),(s1a),(s1b)].
While the latter will receive $1/M_0$ corrections, in the former
baryons do not appear in the loop therefore the tadpole contribution is 
the same as in the relativistic limit. Notice also that
sunset diagrams are proportional to the axial couplings $F$ and $D$, which are not present in the tadpoles.
For this reason the Ademollo-Gatto theorem holds independently on each of the 
two sets of diagrams.

The results we found for $\op(p^2)$ corrections, according to the notation of eq.~(\ref{eq:not}), read
\bea \label{eq:resop2}
\alpha^{(2)}_{\Sigma^- n}&=&-\frac38 (H_{\eta K}+ H_{\pi K})-\frac98(D-F)^2 (H_{\eta K}+H_{\pi K})+	D^2 H_{\pi K}\,, \nn \\
\alpha^{(2)}_{\Lambda p}&=&-\frac38 (H_{\eta K}+ H_{\pi K})-\frac18(D+3F)^2 (H_{\eta K}+H_{\pi K}) -D^2 H_{\pi K}\,, \nn \\
\alpha^{(2)}_{\Xi^- \Lambda}&=&-\frac38 (H_{\eta K}+ H_{\pi K})-\frac18(D-3F)^2 (H_{\eta K}+H_{\pi K})-D^2 H_{\pi K}\,, \nn \\
\alpha^{(2)}_{\Xi^- \Sigma^0}&=&-\frac38 (H_{\eta K}+ H_{\pi K})-\frac98(D+F)^2 (H_{\eta K}+H_{\pi K})+D^2 H_{\pi K}\,.
\eea
The  $\op(p^2)$ function $H_{pq}$ is defined as
\beq
H_{pq}\doteq\frac{1}{(4\pi f)^2}\l[m_p^2+m_q^2-2\frac{m_p^2m_q^2}{m_q^2-m_p^2}\log{\l(\frac{m_q^2}{m_p^2}\r)}\r]\,,
\eeq
and satisfies the Ademollo-Gatto theorem.

In eq.~(\ref{eq:resop2}) the contributions independent of $F$ and $D$ come from tadpoles while 
the rest come from sunset diagrams. It is interesting to notice that the tadpole contributions is
universal (the same for all the channels) and is actually the \emph{same} as in
the $K^0\to\pi^-$ \vff:
\beq
-\frac{3}{8}(H_{\eta K}+ H_{\pi K})= -0.023\,.
\eeq
The sunset contributions, on the other hand, are different in each channel, and can have either signs.
Moreover, these contributions receive important $1/M_0$ and $\delta M_B$ corrections as shown in the 
next sections.

The results in eq.~(\ref{eq:resop2}) agree with those of Krause \cite{Krause} and
Kaiser \cite{Kaiser}. However they do not agree with those of Anderson and Luty \cite{al}, 
which seem to fail the overall sign for the tadpole contributions.
The total $\op(p^2)$ corrections, using physical values 
for masses and couplings (see table~\ref{tab:values}),
are reported in table~\ref{tab:res8}.
\begin{table}
\begin{center}
\begin{tabular}{| c | c  || c | c || c | c |}
\hline 
$m_\pi$ & 0.138& $M_N$ & 0.939& $M_{\Delta_0}$ & 1.232\\
$m_K$ & 0.496& $M_\Sigma$ & 1.193& $M_{\Delta_1}$ & 1.384\\
$m_\eta$ & 0.548& $M_\Lambda$ & 1.116& $M_{\Delta_2}$ & 1.533\\
$f$ & 0.132& $M_\Xi$ & 1.318& $M_{\Delta_3}$ & 1.672\\
$D$ & 0.804& $M_0$ & 1.151& $\Delta$ & 0.231\\
$F$ & 0.463& $\cC$ & 1.6& $$ & \\
\hline 
\end{tabular}
\end{center}
\caption{{\it Numerical values for masses and couplings used in the text 
(dimensionful quantities are in GeV).}}
\label{tab:values}
\end{table}

\subsection{$\op(1/M_0)$ relativistic corrections}
\label{sec:opM}
The $\op(1/M_0)$ corrections due to the non-relativistic expansion are produced by the terms
in eq.~(\ref{eq:LB2M}). There are corrections to the propagators, the diagrams (z3a), (s3g), (s3h) in
figure~\ref{fig:graph8}, to the strong vertices [(z3b),(z3c),(s3a),(s3b),(s3c),(s3d)] and to the vector current
[(s3e),(s3f)]. All these contributions come from sunset diagrams, are $F$ and $D$ dependent and provide
the relativistic corrections to the sunset contributions of eq.~(\ref{eq:resop2}).
They give the following contributions to the \vff:
\bea
\alpha^{(1/M)}_{\Sigma^- n}&=&-\frac98(D-F)^2 ( H'_{\eta K}+ H'_{\pi K})+ D^2  H'_{\pi K}\,, \nn \\
\alpha^{(1/M)}_{\Lambda p}&=&-\frac18(D+3F)^2 ( H'_{\eta K}+ H'_{\pi K})- D^2  H'_{\pi K}\,, \nn \\
\alpha^{(1/M)}_{\Xi^- \Lambda}&=&-\frac18(D-3F)^2 ( H'_{\eta K}+ H'_{\pi K})-D^2  H'_{\pi K}\,, \nn \\
\alpha^{(1/M)}_{\Xi^- \Sigma^0}&=&-\frac98(D+F)^2 ( H'_{\eta K}+ H'_{\pi K})+D^2  H'_{\pi K}\,,
\label{eq:resopM}
\eea
where
\beq
H'_{p,q}\doteq-\frac{2\pi}{3 (4\pi f)^2 M_0} \frac{(m_p-m_q)^2}{m_p+m_q}\l ( m_p^2+3m_pm_q+m_q^2\r)\,,
\eeq
satisfies the Ademollo-Gatto theorem.
The contributions of eqs.~(\ref{eq:resopM}) were not considered in refs.~\cite{al,Kaiser}.
They agree with the $1/M_0$ expansion of the relativistic result of ref.~\cite{Krause}.
This is due to the fact that the relativistic corrections of eq.~(\ref{eq:resopM}), at this order, 
are non-analytic in the quark masses, thus according to the discussion in \cite{bl}, do not break the chiral expansion.

The physical values for the corrections in eq.~(\ref{eq:resopM}) are reported 
in table~\ref{tab:res8}. They are important,
as already noticed in \cite{Krause}, and tend to cancel the $F$ and $D$ $\op(p^2)$ contributions.

\begin{figure}[t]
\psfrag{X}{\hspace{-24pt} $M_0$~(GeV)} 
\psfrag{Y}{\hspace{-24pt} $\frac{f_1(0)}{f_1^{SU(3)}(0)}-1$}
\bc \epsfig{file=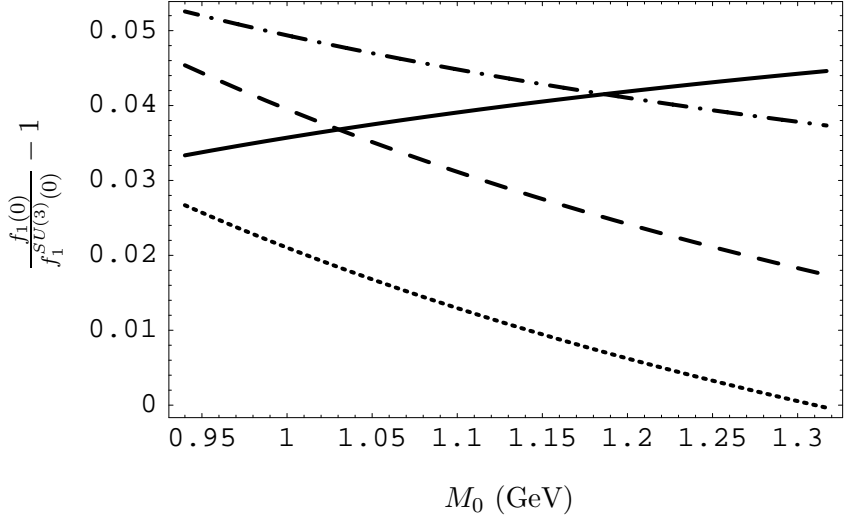,width=0.65\textwidth} \ec 
\vspace{-0.6cm}
\caption{{\it Plot of the SU(3)-breaking corrections to the \vff\ as functions of $M_0$. 
Solid, dashed, dot-dashed and dotted curves represent
the channels $\Sigma^-\to n$, $\Lambda\to p$, $\Xi^-\to\Lambda$ and $\Xi^-\to\Sigma^0$ respectively.}}
\label{fig:M0dep}
\end{figure}

Moreover, these corrections contain the new parameter $M_0$, which is actually defined as the baryon mass in the 
chiral limit. At this order, however, it should be safe to choose any physical baryon mass,
as long as the perturbative expansion holds, since the difference would be of $\op(p^5)$.
By varying $M_0$ one can check this assumption and infer on the importance of higher order effects.
The result is illustrated in figure~\ref{fig:M0dep}, which shows that the dependence on $M_0$ is quite strong.
In the plot is actually reported the dependence on $M_0$ of the full SU(3)-breaking corrections up to $\op(p^4)$.
This fact was somewhat expected since, as observed also in the mesonic case, the chiral expansion
with three flavors converges very slowly and contributions as large as the leading one can be expected
from the resummation of higher order corrections.
\subsection{$\op(p^3)$ corrections}
\label{sec:op3oct}
$\op(p^3)$ corrections to the \vff\ are produced by one-loop diagrams
via the insertion of one $\op(p^2)$ operator. There are many $\op(p^2)$ operators 
with unknown LECs in the \HBChPT. However, only those generating
baryon mass shifts and reported explicitly in eq.~(\ref{eq:LB2p3}) give non-vanishing
contributions to the \vff. This allows to study also $\op(p^3)$ corrections without
the uncertainties coming from higher order LECs.

In order to calculate these contributions to the \vff\ one can just
add mass-shift effects to the baryon propagators in the $\op(p^2)$ diagrams.
Note that tadpole diagrams do not contribute because baryons do not enter
the loop (external lines are taken on-shell at the shifted mass value). 
Therefore, also the $\op(p^3)$ corrections, as the $1/M_0$ ones, will be proportional to the axial couplings 
$D$ and $F$. There is however a subtlety. At this order, indeed, the incoming and 
the outcoming baryons will no longer be degenerate in mass. In particular
this means that, at $q^2=0$, $q^\mu\neq0$ and a non vanishing momentum has to be 
injected into the loop. We can parametrize on-shell momenta as follows:
\bea \label{eq:qshift}
&&p_1^\mu=M_1\, v^\mu\,,\qquad p_2^\mu=M_1\, v^\mu-q^\mu\,, \qquad M_{1(2)}=M_0+\delta M_{1(2)} \nn \\
&&q^2=0\,,\qquad p_2^2=M_1^2-2M_1\,v\cdot q=M_2^2\,,\nn \\ 
&\Rightarrow&  v\cdot q=\frac{M_1^2-M_2^2}{2M_1}\simeq (\delta M_1-\delta M_2)+\op(\delta M)^2\,.
\eea
Since $\delta M_{1(2)}\sim \op(p^2)$ also $v\cdot q\sim \op(p^2)$ and it can be expanded as a mass
insertion. The corresponding contribution is represented as a mass insertion
on external lines in the graphs (z2), (s2a) and (s2b) in figure~\ref{fig:graph8}.

The explicit expression for the $\op(p^3)$ chiral corrections is rather lengthy and
is reported in appendix~\ref{sec:app1}. 
Notice that, because of the Ademollo-Gatto
theorem only the differences of baryon masses appear ($\delta M_1-\delta M_2=M_1-M_2$),
and the result does not depend on the choice of $M_0$.

The same contributions were calculated in ref.~\cite{al}. 
However we do not agree with that result. Ademollo-Gatto theorem
can be checked explicitly from our eqs.~(\ref{eq:resp31})-(\ref{eq:resp34}).
In \cite{al} the authors declare to have made the same check too,
unfortunately only the formul\ae\ for $m_\pi=0$ are given in \cite{al}.
Numerically our results give smaller SU(3)-breaking corrections 
(see table~\ref{tab:res8}). However they still remain important with respect to
the leading corrections and cannot be neglected.

Note that these contributions do not receive $1/M_0$ relativistic corrections 
at this order in the chiral expansion. 
These contributions, indeed, would be $\op(p^5)$, which we did not consider here.

\subsection{Final results for octet contributions}
\label{sec:finoct}
In table~\ref{tab:res8} we showed the numerical estimates for the sum 
of all 1-loop contributions up to $\op(p^4)$ using the physical values
for masses and couplings of table~\ref{tab:values}.
The final results are slightly smaller
than previously claimed\footnote{Note that in \cite{al} authors
used chiral limit values for $F$ and $D$, which are sensibly smaller
than the physical ones. We preferred to use the latter which are better
known. The difference, assuming the chiral expansion holds, is an higher
order effect.} in ref.~\cite{al}. However they are still large compared
to what expected by CKM unitarity \cite{csw}, and opposite in signs
with respect to quark model estimates \cite{qmodel}. 
The explicit calculation of the \vff\ shows that $\op(p^3)$ and $1/M_0$ corrections
are not small and the expansion converge very slowly.
By varying $M_0$ (see figure~\ref{fig:M0dep}) and the parameters $D$ and $F$ between their
physical value (table~\ref{tab:values}) and what is expected to be their value 
in the chiral limit (see e.g. \cite{jm1}), SU(3) corrections vary substantially.
This is manly due to the fact that the dependence on $D$ and $F$
is quadratic and the various corrections in table~\ref{tab:res8}
tend to cancel each other. 
Going beyond the $\op(p^4)$ calculation would be challenging and useless
since unknown LECs would appear. The impact of higher $1/M_0$ corrections,
on the other hand, could be studied by calculating the amplitude
within a relativistic framework (see e.g. \cite{bl}).
Note also that lattice QCD \cite{hyplat}
and quark models calculations \cite{qmodel}
suggest that local contributions (which would start at $\op(p^5)$
in the chiral expansion) give a negative contribution.

Another issue is represented by contributions from the decuplet states.
In the calculation above decuplet states are taken to be integrated out,
and their contribution reabsorbed into the parameters of the chiral
Lagrangian. Decuplet degrees of freedom, however, are not so heavy to be
safely considered frozen and non-analytic contributions can be important.
We will study their effects in the next section.
For all these reasons, the uncertainty in the SU(3)-breaking corrections
of table~\ref{tab:res8} should be taken of order one. 

As last remark we must say that the lack of convergence of the $\op(p^4)$ Lagrangian
does not appear peculiar of the \vff\ only, 
but of the 3-flavors \HBChPT\ expansion itself. In view of these results, we think that 
it is dangerous to trust one-loop calculations, especially if $1/M_0$ and 
$\op(p^3)$ corrections are not taken into account. For other quantities, indeed,
the lack of convergence can hide itself behind the ignorance of the LECs.

\begin{table}
\begin{center}
\begin{tabular}{ c | c |  c  c  c | c }
\hline 
Decay & $\alpha^{(1)}$ & $\alpha^{(2)}$($\times 10^2$) & 
	$\alpha^{(3)}$($\times 10^2$) & $\alpha^{(1/M)}$($\times 10^2$) & All ($\times 10^2$) \\ \hline 
$\Sigma^-\to n$ & $-1$ & $+0.7$  & $+6.5$  & $-3.2$  & $+4.1$  \\ 
$\Lambda\to p$ & $-\sqrt{3/2}$ & $-9.5$  & $+4.3$  & $+8.0$  & $+2.7$  \\ 
$\Xi^-\to \Lambda$ & $\sqrt{3/2}$ & $-6.2$  & $+6.2$  & $+4.3$  & $+4.3$  \\ 
$\Xi^-\to \Sigma^0$ & $1/\sqrt2$ & $-9.2$  & $+2.4$  & $+7.7$  & $+0.9$  \\ 
\hline 
\end{tabular}
\end{center}
\caption{{\it Numerical estimates of the chiral corrections 
(All $=\alpha^{(2)}+\alpha^{(3)}+\alpha_{1/M}$) to the \vff. 
The physical values used for the parameters are reported in table~\ref{tab:values}.}}
\label{tab:res8}
\end{table}

\section{Dynamical Decuplet}
\label{sec:dec}
The calculations made so far rely on the assumptions that all other hadronic states
can be safely integrated out. This is a good approximation if 
the relevant momenta are much smaller than the
scale corresponding to higher hadronic states. In particular we need $p\sim m_{\pi,K}\ll \Delta$,
where $\Delta$ is the mass shift between octet baryons and the lightest
excitations. In QCD, these excitations are represented by the decuplet states,
which unfortunately are rather light $\Delta\sim230$~MeV. This means that 
these states can give dangerous non-analytic contributions
that cannot be reabsorbed into local counterterms. Whether these contributions
are actually important depends both on how the decuplet couples to the other fields
and on the specific quantity under study. In order to make this statement
more quantitative (H)B\ChPT\ has been extended also to these degrees of freedom \cite{jm2}.
An issue that arises in embedding the decuplet within the framework of \HBChPT,
is how to deal with the decuplet mass shift scale $\Delta$. This
scale, indeed, is neither large enough to be integrated out,
nor is a chiral parameter, which could be tuned to be arbitrary small to make
the expansion hold. In \cite{hhk} a $phenomenological$ expansion was proposed
where $\Delta\sim m_\pi$ is treated as a small parameter of $\op(p)$.
Within this framework the decuplet contributions to the \HBChPT\ Lagrangian 
can be organized as follows:
\beq
\cL_{10}=\cL^{(1)}_\Delta+\cL^{(2)}_\Delta+\cL^{(3)}_\Delta+\dots
\eeq
where the leading order terms read
\bea \label{eq:L1D}
&&\cL^{(1)}_\Delta=-i\ov T^\mu v^\nu  D_\nu T_\mu +\Delta\, \ov T^\mu T_\mu
	+\cC \l ( \ov T^\mu A_\mu B+ \ov B A_\mu T^\mu \r ) +
	2\, \cH\, \ov T^\mu S^\nu A_\nu T_\mu\,, \\	
&& \hspace{120pt} D_\mu T_\nu = \de_\mu T_\nu-iV_\mu T_\nu\,. \nn
\eea
In eq.~(\ref{eq:L1D}) flavor indices are not explicitly shown and
$T_\mu$ is the light component of the spin-3/2 Rarita-Schwinger field 
associated to the decuplet, the propagator of which (in $d$-dimensions) reads:
\beq
\frac{i}{k v - \Delta+i\eps} \l ( v_\mu v_\nu-g_{\mu\nu}-4\,\frac{d-3}{d-1} S_\mu S_\nu \r)\,.
\eeq
In eq.~(\ref{eq:L1D}), $\cC$ and $\cH$ are the effective couplings associated 
to the decuplet-octet-meson and the decuplet-decuplet-meson vertices, respectively. 
$\cC$ can be extracted from $\Delta\to\pi N$ decays (\cite{jm2}). 
$\cH$, which is poorly known, does not
contribute, however, to the hyperon \vff\ at the order we are working.

\begin{figure}[t]
\bc \epsfig{file=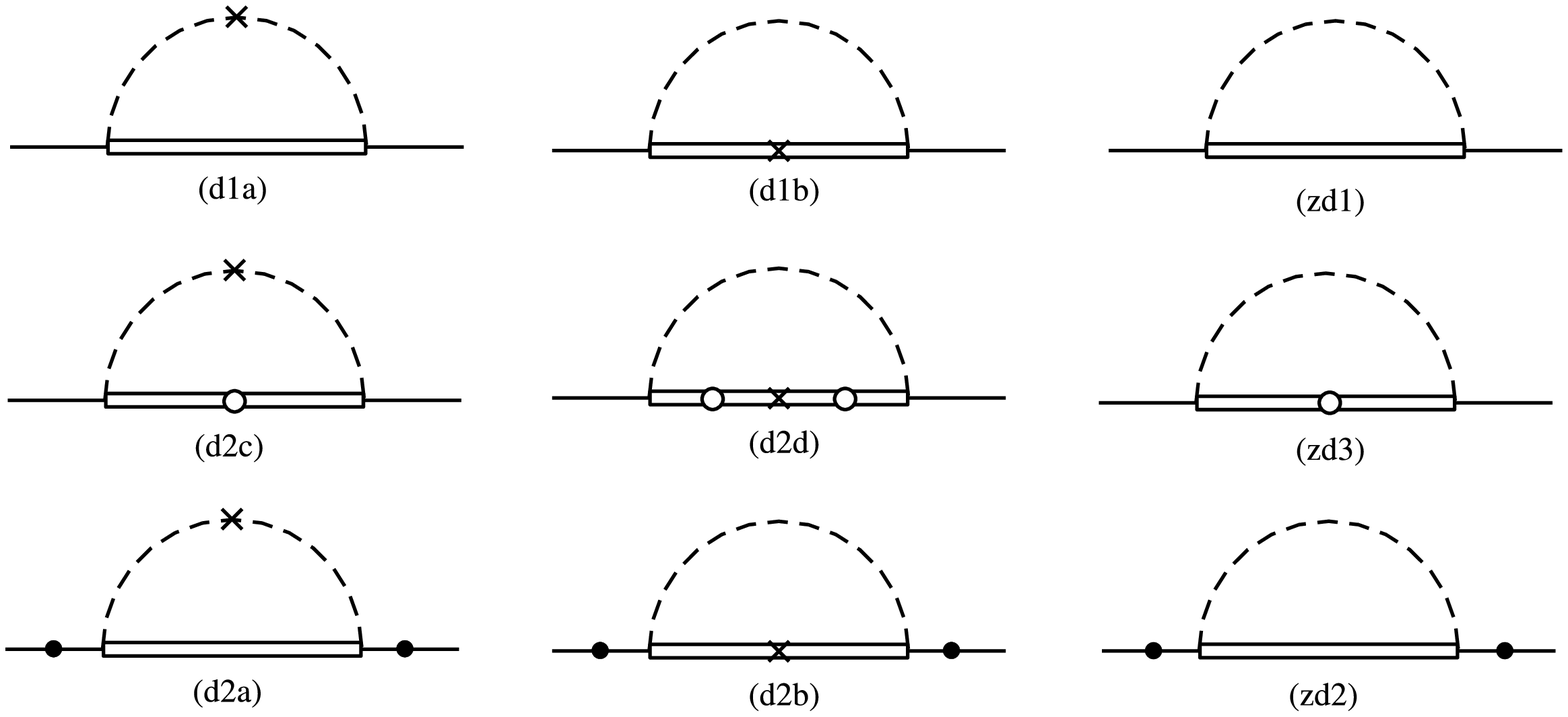,width=0.85\textwidth} \ec 
\caption{{\it Feynman diagrams for decuplet contributions to hyperon \vff.
Notations are the same as in figure~\ref{fig:graph8} with
double line and empty circles representing respectively decuplet propagators and
SU(3)-breaking decuplet mass-shift operators ($\delta M_\Delta$ in the text).}}
\label{fig:graphs10}
\end{figure}
Analogously to the octet case, we can parametrize the SU(3)-breaking corrections
due to the decuplet as follows:
\beq
\alpha^{(1)} \l[\beta^{(2)} + \l(\beta^{(3)}+\beta^{(1/M)}\r)+\dots \r] \,.
\eeq
The lowest order decuplet corrections $\beta^{(2)}$ to the \vff\
are $\op(p^2)$ with respect to the tree-level amplitudes $\alpha^{(1)}$ 
and are represented by the diagrams [(d1a),(d1b),(zd1)] 
in figure~\ref{fig:graphs10}. All the contributions come from sunset
diagrams and, as in the octet case, they sum up to give finite results,
according to the Ademollo-Gatto theorem.
We find:
\bea \label{eq:resop2d}
\beta^{(2)}_{\Sigma^- n}&=&\frac23 \cC^2 \l (G_{\eta K}-\frac38 H_{\eta K}\r)
	+\frac43 \cC^2 \l(G_{\pi K}-\frac38 H_{\pi K} \r ) \,, \nn \\
\beta^{(2)}_{\Lambda p}&=&-\frac23 \cC^2 \l (G_{\pi K}-\frac38 H_{\pi K} \r )\,, \nn \\
\beta^{(2)}_{\Xi^- \Lambda}&=&\frac23\cC^2 \l ( G_{\eta K}-\frac38 H_{\eta K} \r )\,, \nn \\
\beta^{(2)}_{\Xi^- \Sigma^0}&=&-\frac43 \cC^2 \l (G_{\eta K}-\frac38 H_{\eta K}\r)
	-\frac23 \cC^2 \l( G_{\pi K}-\frac38 H_{\pi K} \r )\,, \nn
\eea
where
\bea
G_{p q}&\equiv&
\frac{\Delta^2}{(4\pi f)^2} \l [ 1- \frac{3m_p^2+3m_q^2-4\Delta^2}{2(m_p^2-m_q^2)} \log \l(\frac{m_p^2}{m_q^2}\r)\r]+ \\
&&+\frac{\Delta}{(4\pi f)^2} \l[ 
\frac{m_p^2+3m_q^2-4\Delta^2}{m_p^2-m_q^2}\sqrt{m_p^2-\Delta^2}\arccos\l(\frac{\Delta}{m_p}\r)+\r. \nn \\
&&\qquad \l.+\frac{m_q^2+3m_p^2-4\Delta^2}{m_q^2-m_p^2}\sqrt{m_q^2-\Delta^2}\arccos\l(\frac{\Delta}{m_q}\r)\r]\nn \,,
\eea
satisfies the Ademollo-Gatto theorem.
 
\begin{figure}[t]
\psfrag{X}{\hspace{-24pt} $\Delta$~(GeV)}
\psfrag{Y}{\hspace{-24pt} $\beta^{(2)}_{\Sigma^{-} n}$}
\bc \epsfig{file=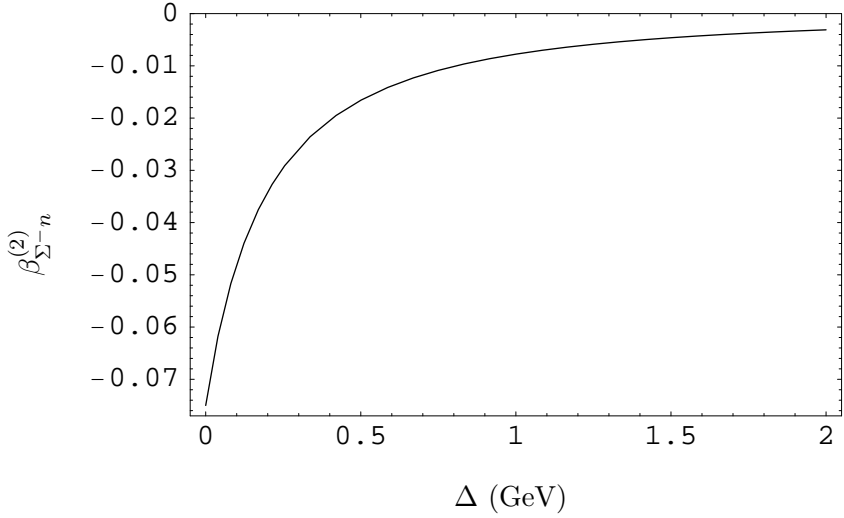,width=0.65\textwidth} \ec 
\caption{$\op(p^2)$ decuplet contributions as a function of $\Delta$.}
\label{fig:decoupling}
\end{figure}

As expected, in the limit $\Delta\to \infty$ the decuplet decouples and its 
contribution goes to zero. In fact,
\beq
\lim_{\Delta\to\infty} \l(G_{pq}-\frac38 H_{pq}\r)=\op\l(\frac{p^4}{\Delta^2}\r)\,,
\eeq
and the contributions of eq.~(\ref{eq:resop2d}) can be reabsorbed into
local $\op(p^5)$ counterterms.
As an example, in figure~\ref{fig:decoupling} we show the plot of
the $\op(p^2)$ decuplet corrections to the $\Sigma^-\to n$ \vff\ as a function
of $\Delta$.

The resulting values for these decuplet contributions are given in table~\ref{tab:res10}.
They are sizable, however, in order to have a reliable
estimate, we need also to include the subleading $\op(p^3)$ corrections.

As for the octet case the $\op(p^3)$ corrections come from the
insertion of $\op(p^2)$ mass-shift operators in one-loop diagrams. In this case, we have two contributions,
coming from the insertion of decuplet and octet mass shifts (diagrams (d2c), (d2d), (zd3)
and (d2a), (d2b), (zd2) in figure~\ref{fig:graphs10}, respectively):
\beq \label{eq:beta3tot}
\beta^{(3)}=\beta^{(3)}(\delta M_\Delta)+\beta^{(3)}(\delta M_B)\,.
\eeq

The corrections due to the decuplet mass splitting read:
\bea \label{eq:resop3d1}
\beta^{(3)}_{\Sigma^- n}(\delta M_\Delta)&=&\cC^2 \l [ \frac23 \delta M_{\Delta_1} G'_{\eta K}
	+\frac43\l( \frac{4\,\delta M_{\Delta_0}-\delta M_{\Delta_1}}{3}\r) G'_{\pi K}	\r ] \,, \nn \\
\beta^{(3)}_{\Lambda p}(\delta M_\Delta)&=&-\cC^2 \l[\frac23 \delta M_{\Delta_1} G'_{\pi K}\r]\,, \nn \\
\beta^{(3)}_{\Xi^- \Lambda}(\delta M_\Delta)&=&\cC^2 \l [ \frac23 \delta M_{\Delta_2} G'_{\eta K}
	 +\frac23 \l(\delta M_{\Delta_1}-\delta M_{\Delta_2}\r) G'_{\pi K} \r ]\,, \nn \\
\beta^{(3)}_{\Xi^- \Sigma^0}(\delta M_\Delta)&=&- \cC^2 \l [\frac43\l(\frac{\delta M_{\Delta_1}+\delta M_{\Delta_2}}{2}\r) G'_{\eta K}
	+\frac23 \l(\frac{2\,\delta M_{\Delta_1}+\delta M_{\Delta_2}}{3}\r) G'_{\pi K} \r ]\,,
\eea
where
\beq
G'_{pq}=\frac{\de G_{pq}}{\de \Delta}\,,
\eeq
obviously satisfies the Ademollo-Gatto theorem.
$\delta M_{\Delta_i}$ are the SU(3)-breaking mass shifts with respect to 
the scale $\Delta$, i.e.:
\beq
\delta M_{\Delta_i}\equiv M_{\Delta_i}-(M_0+\Delta)\,,
\eeq
and $\Delta_{0,1,2,3}$ correspond to the decuplet states $\Delta(1232)$, $\Sigma(1385)$,
$\Xi(1530)$ and $\Omega^-$ respectively.

The corrections in eq.~(\ref{eq:resop3d1}) are numerically small 
(see table~\ref{tab:res10}) and, as the $\op(p^2)$ ones, vanish 
in the limit $\Delta\to\infty$.
Note, however, that the contributions with small corrections in table~\ref{tab:res10} (i.e. those below $\sim$1\%)
result by accidental cancellations, so that their values strongly depend
on the specific choice of masses and couplings.

Finally, we have the corrections from octet mass shifts. As shown in
figure~\ref{fig:graphs10} the only contributions come from mass insertions
on external legs, which shift the transferred momentum as discussed in
section~\ref{sec:op3oct}. The explicit expression for the corresponding
$\beta^{(3)}(\delta M_B)$ contributions  are lengthy and 
reported in appendix~\ref{sec:app2}.
The Ademollo-Gatto theorem has been checked as well as the decoupling limit.

At a first sight, both the $\beta^{(3)}(\delta M_\Delta)$ 
and $\beta^{(3)}(\delta M_B)$ contributions depend on the choice of $M_0$ 
(via the mass shifts $\delta M_\Delta$ and $\delta M_B$). However,
it can be checked that this dependence cancel in the sum $\beta^{(3)}$ 
of eq.~(\ref{eq:beta3tot}), and the total $\op(p^3)$ corrections are 
independent of the explicit choice of $M_0$ 
as in the octet contribution $\alpha^{(3)}$.

Evaluating the $\beta^{(3)}(\delta M_B)$ contributions 
numerically (see table~\ref{tab:res10})
we can notice that, in general, these corrections are huge. 
These results seem to signal a breaking of the perturbative expansion.
\begin{table}
\begin{center}
\begin{tabular}{ c | c   c  c }
\hline 
Decay &  $\beta^{(2)}$ & $\beta^{(3)}(\delta M_\Delta)$ & $\beta^{(3)}(\delta M_B)$  \\ \hline 
$\Sigma^-\to n$ & $-3.1$  & $-1.8$  & $+38.1$    \\ 
$\Lambda\to p$ &  $+1.5$  & $-0.01$  & $+6.9$    \\ 
$\Xi^-\to \Lambda$ & $-0.05$  & $-0.6$  & $+18.4$  \\ 
$\Xi^-\to \Sigma^0$ &  $+1.6$  & $-0.2$  & $-1.3$  \\ 
\hline 
\end{tabular}
\end{center}
\caption{{\it Decuplet contributions to the chiral corrections ($\times 10^2$) to the \vff.
Numerical values for the parameters are spelled out in table~\ref{tab:values}.}}
\label{tab:res10}
\end{table}

We would like to stress again that the estimates in tables~\ref{tab:res8} 
and \ref{tab:res10} do not rely on unknown higher order LECs. They
involve just the well known tree level parameters: masses, decay constants
and 3-particle couplings (table~\ref{tab:values}). As long as the perturbative expansion holds, 
chiral corrections to these parameters must
be considered of higher order, which justifies the use of the physical 
values.
\subsection{Discussion}
\label{sec:disdec}
The results obtained for the decuplet contributions deserve more discussion.
Already in the pure octet case we observed a slow convergence of
the chiral expansion. However, with the inclusion of decuplet states, 
it seems that the perturbative series breaks completely. 
Based on the fact that the decuplet couples to mesons stronger 
than the octet ($\cC^2/D^2\sim 4$) 
we could have expected a large contribution from the former.
This however cannot explain the large $\beta^{(3)}$ contributions with respect
to $\beta^{(2)}$.

Notice also that $\op(1/M_0)$ corrections cannot cure the expansion since, at
this order, they can only affect $\beta^{(2)}$. Relativistic corrections 
to $\beta^{(3)}$ are indeed of higher order in our power counting.

An alternative approach which could allow to resum all the $1/M_0$
corrections is the use of the relativistic formulations for Baryon
Chiral Perturbation Theory \cite{bl}. These expansions are based on the use 
of modified regularizations, which allow to reabsorb into the LECs 
the terms that would break the power counting.
In the particular case of the hyperon \vff\ discussed here,
this implementation should work in a peculiar way since all
one loop contributions are finite, thus independent on the
regularization scheme. Although these approaches would improve
the calculation made here and probably help in reducing the anomalous
decuplet contributions, is unlike that these could heal the problem completely,
since extraordinary fine tuned relativistic corrections would be needed.

Moreover, there are clues indicating that the observed
breaking of the chiral expansion is more related to the flavor
power counting rather than to the relativistic corrections.
In the standard power counting, indeed, the quark masses are $\op(p^2)$
so that meson masses ($m_{\pi,K\dots}$) are $\op(p)$ while 
baryon splittings ($\delta M_B$, $\delta M_\Delta$) are $\op(p^2)$.
Following \cite{hhk} we treated the octet-decuplet mass shift 
$\Delta$ as a phenomenological parameter of $\op(p)$ since numerically
$\Delta\sim m_\pi$. However, while SU(2)-breaking baryon mass splittings
are suppressed with respect to $\Delta$, the SU(3) ones are not.
Note that if $\Delta$ is considered to be $\op(p^2)$, the chiral 
corrections get even worse, while considering $\Delta\sim\op(1)$ would 
correspond to integrating out the decuplet.
It does not seem that the convergence problem can be cured in this way and 
doubts arise on the reliability of (H)B\ChPT\ with three flavors\footnote{The problem
of the convergence of the (H)B\ChPT\ with three flavors has already been discussed 
in other cases (see e.g. \cite{bchpt3}). However, \vff\ allow for a more 
reliable test of the perturbative expansion because, unlike other quantities, 
do not depend on unknown LECs at this order.}.

\section{Conclusions}
\label{sec:concl}
Recent progresses in the extraction of $V_{us}$ from hyperon semileptonic decays \cite{csw}
made the estimation of the SU(3)-breaking effects for the corresponding vector
form factors a central issue. 
Numerical studies \cite{kl3hyplat,hyplat} demonstrate the possibility to estimate such corrections.
However, current simulations are not yet able to catch the non-analytic
contributions of the meson loops, which in these processes are the
dominant ones because of the Ademollo-Gatto theorem.
In this paper we estimated these contributions in the framework
of Heavy Baryon Chiral Perturbation Theory. We performed a full
$\op(p^4)$ calculation including relativistic ($1/M_0$) corrections,
extending and correcting previous analysis \cite{Krause,al,Kaiser}.
An important fact is that the Ademollo-Gatto theorem guarantees for
these quantities the absence of local contributions at this order, 
therefore the final estimates are free from unknown parameters.
We show that the would-be subleading $\op(p^4)$ contributions are important
and signal a poor convergence of the chiral expansion. 
This might not be so bad, anyway. We know that the three-flavor
chiral expansion is slowly converging, and also in the case of
kaon semileptonic decays \cite{lr,kl3}, important contributions come from
the subleading local contributions, which can be estimated, for instance,
using Lattice QCD.

The chiral expansion for baryons, however, presents also other complications.
In particular the calculations made above rely on the fact that heavier 
hadronic states could be safely decoupled. This approximation may however be
not so good for the decuplet states, which are only slightly heavier than the octet
baryons. These states have been implemented into the (H)B$\chi$PT framework in \cite{jm2,hhk}.
We thus evaluated the corresponding contributions to the \vff\
to $\op(p^4)$. We found that $\op(p^3)$ decuplet contributions, in general, may be important,
in agreement with analysis made for other quantities (see e.g. \cite{jm2}). The $\op(p^4)$ contributions, however,
are anomalously huge, signaling a breaking of the perturbative expansion.
These results arise doubts on the consistency of the chiral expansion with
the decuplet, at least for the 3-flavors case.

Recently, several progress have been made towards a consistent relativistic
formulation of the chiral expansion for baryons \cite{bl}. Such approaches could
in principle remove part of the uncertainties connected to the $1/M_0$ expansion
and its slow convergence. It is unlike, however, that they could solve the
issues connected to the decuplet states, which seem more related to the structure of the power
counting of the chiral flavor parameters.

%
\section*{Acknowledgments}
We thank D.~Guadagnoli, G.~Isidori, V.Lubicz, G.~Martinelli and S.~Simula for 
useful discussions and comments on the manuscript.
\appendix
\section{Octet $\op(p^3)$ contributions}\label{sec:app1}
We present here the explicit expressions for the $\op(p^3)$ corrections $\alpha^{(3)}$
due to octet baryon contributions to hyperon \vff:

\footnotesize
\bea \label{eq:resp31}
&&\hspace{-12pt}\alpha^{(3)}_{\Sigma^- n}=\nn \\ &&\frac{1}{144\,\pi \,f^2\,
    {\left( m_\eta + m_K \right) }^2\,
    {\left( m_K + m_\pi \right) }^2}\Bigl\{
	3\,F^2\,\Bigl [ 6\,m_\eta^3\,
        {\left( m_K + m_\pi \right) }^2 + 
       m_\eta^2\,\left( m_K^3 + 2\,m_K^2\,m_\pi + 
          m_K\,m_\pi^2 - 10\,m_\pi^3 \right) \nn \\ && + 
       2\,m_\eta\,m_K\,
        \left( m_K^3 + 2\,m_K^2\,m_\pi + 
          m_K\,m_\pi^2 - 10\,m_\pi^3 \right)  + 
       m_K^2\,\left( m_K^3 + 2\,m_K^2\,m_\pi + 
          m_K\,m_\pi^2 - 10\,m_\pi^3 \right)  \Bigr ] \,
     \left( M_N - M_\Sigma \right)   \nn \\ && -
    3\,D\,F\,\Bigl [ 4\,m_\eta^3\,
        {\left( m_K + m_\pi \right) }^2\,
        \left( M_N - M_\Sigma \right)  + 
       m_\eta^2\,\left( m_K^3\,
           \left( 3\,M_\Lambda + 2\,M_N - 5\,M_\Sigma \right)  - 
          m_K^2\,m_\pi\,
           \left( 3\,M_\Lambda - 4\,M_N + M_\Sigma \right) \r.  \nn \\ && \l.+ 
          m_K\,m_\pi^2\,
           \left( -3\,M_\Lambda + 2\,M_N + M_\Sigma \right)  + 
          3\,m_\pi^3\,\left( M_\Lambda - 4\,M_N + 3\,M_\Sigma
             \right)  \right)  + 2\,m_\eta\,m_K\,
        \left( m_K^3\,\left( 3\,M_\Lambda + 2\,M_N - 
             5\,M_\Sigma \right)  \r. \nn \\ && \l. - 
          m_K^2\,m_\pi\,
           \left( 3\,M_\Lambda - 4\,M_N + M_\Sigma \right)  + 
          m_K\,m_\pi^2\,
           \left( -3\,M_\Lambda + 2\,M_N + M_\Sigma \right)  + 
          3\,m_\pi^3\,\left( M_\Lambda - 4\,M_N + 3\,M_\Sigma
             \right)  \right)   \nn \\ && + m_K^2\,
        \left( m_K^3\,\left( 3\,M_\Lambda + 2\,M_N - 
             5\,M_\Sigma \right)  - 
          m_K^2\,m_\pi\,
           \left( 3\,M_\Lambda - 4\,M_N + M_\Sigma \right)  + 
          m_K\,m_\pi^2\,
           \left( -3\,M_\Lambda + 2\,M_N + M_\Sigma \right)  \r. \nn \\ &&\l. + 
          3\,m_\pi^3\,\left( M_\Lambda - 4\,M_N + 3\,M_\Sigma
             \right)  \right)  \Bigr]  -
    {D}^2\,\Bigl [ 6\,m_\eta^3\,
        {\left( m_K + m_\pi \right) }^2\,
        \left( M_N - M_\Sigma \right)  + 
       m_\eta^2\,\left( m_K^3\,
           \left( 3\,M_\Lambda + M_N - 4\,M_\Sigma \right)  \r. \nn \\ && \l.+ 
          m_K^2\,m_\pi\,
           \left( -3\,M_\Lambda + 2\,M_N + M_\Sigma \right)  + 
          m_K\,m_\pi^2\,
           \left( -3\,M_\Lambda + M_N + 2\,M_\Sigma \right)  + 
          m_\pi^3\,\left( 3\,M_\Lambda - 10\,M_N + 
             7\,M_\Sigma \right)  \right) \nn \\ && + 
       2\,m_\eta\,m_K\,
        \left( m_K^3\,\left( 3\,M_\Lambda + M_N - 
             4\,M_\Sigma \right)  + 
          m_K^2\,m_\pi\,
           \left( -3\,M_\Lambda + 2\,M_N + M_\Sigma \right)  + 
          m_K\,m_\pi^2\,
           \left( -3\,M_\Lambda + M_N + 2\,M_\Sigma \right) \r. \nn \\ && \l.+ 
          m_\pi^3\,\left( 3\,M_\Lambda - 10\,M_N + 
             7\,M_\Sigma \right)  \right)  + 
       m_K^2\,\left( m_K^3\,
           \left( 3\,M_\Lambda + M_N - 4\,M_\Sigma \right)  + 
          m_K^2\,m_\pi\,
           \left( -3\,M_\Lambda + 2\,M_N + M_\Sigma \right) \r. \nn \\ &&\l.+ 
          m_K\,m_\pi^2\,
           \left( -3\,M_\Lambda + M_N + 2\,M_\Sigma \right)  + 
          m_\pi^3\,\left( 3\,M_\Lambda - 10\,M_N + 
             7\,M_\Sigma \right)  \right)  \Bigr] \Bigr\}\,,
\eea
\bea \label{eq:resp32}
&&\hspace{-12pt}\alpha^{(3)}_{\Lambda p}=\nn \\ &&
\frac{1}{48\,\pi \,f^2\,{\left( m_\eta + m_K \right) }^2\,    {\left( m_K + m_\pi \right) }^2} \Bigl \{
	-3\,F^2 \,
     \Bigl [ 2\,m_\eta^3\,{\left( m_K + m_\pi \right) }^2 - 
       m_\eta^2\,\left( m_K^3 + 2\,m_K^2\,m_\pi + 
          m_K\,m_\pi^2 - 2\,m_\pi^3 \right)  
		  \nn \\ &&
		  - 2\,m_\eta\,m_K\,\left( m_K^3 + 2\,m_K^2\,m_\pi + 
          m_K\,m_\pi^2 - 2\,m_\pi^3 \right)  - 
       m_K^2\,\left( m_K^3 + 2\,m_K^2\,m_\pi + 
          m_K\,m_\pi^2 - 2\,m_\pi^3 \right)  \Bigr ] \,\left( M_{\Lambda} - M_{N} \right) 
		\nn \\ &&
		  +     D\,F\,\Bigl [ 4\,m_\eta^3\,\left( M_{\Lambda} - M_{N} \right) \,
        {\left( m_K + m_\pi \right) }^2 + 
       m_\eta^2\,\left( m_K^3\,
           \left( 5\,M_{\Lambda} - 2\,M_{N} - 3\,M_{\Sigma} \right)  - 
          m_K\,m_\pi^2\,\left( M_{\Lambda} + 2\,M_{N} - 3\,M_{\Sigma} \right)
		  \r. \nn \\ && \l.
              - 3\,m_\pi^3\,\left( 3\,M_{\Lambda} - 4\,M_{N} + M_{\Sigma} \right)  + 
          m_K^2\,m_\pi\,\left( M_{\Lambda} - 4\,M_{N} + 3\,M_{\Sigma} \right) 
          \right)  + 2\,m_\eta\,m_K\,
        \left( m_K^3\,\left( 5\,M_{\Lambda} - 2\,M_{N} - 3\,M_{\Sigma} \right)
		  \r. \nn \\ && \l.
		  - 
          m_K\,m_\pi^2\,\left( M_{\Lambda} + 2\,M_{N} - 3\,M_{\Sigma} \right)
              - 3\,m_\pi^3\,\left( 3\,M_{\Lambda} - 4\,M_{N} + M_{\Sigma} \right)  + 
          m_K^2\,m_\pi\,\left( M_{\Lambda} - 4\,M_{N} + 3\,M_{\Sigma} \right) 
          \right) 
			\nn \\ &&		  
		   + m_K^2\,\left( m_K^3\,
           \left( 5\,M_{\Lambda} - 2\,M_{N} - 3\,M_{\Sigma} \right)  - 
          m_K\,m_\pi^2\,\left( M_{\Lambda} + 2\,M_{N} - 3\,M_{\Sigma} \right)
              - 3\,m_\pi^3\,\left( 3\,M_{\Lambda} - 4\,M_{N} + M_{\Sigma} \right)  
		  \r. \nn \\ && \l.
			  + m_K^2\,m_\pi\,\left( M_{\Lambda} - 4\,M_{N} + 3\,M_{\Sigma} \right) 
          \right)  \Bigr]  + D^2\,
     \Bigl[ 2\,m_\eta^3\,\left( M_{\Lambda} - M_{N} \right) \,
        {\left( m_K + m_\pi \right) }^2 + 
       m_\eta^2\,\left( m_K\,m_\pi^2\,
           \left( 2\,M_{\Lambda} + M_{N} - 3\,M_{\Sigma} \right)  
		  \r. \nn \\ && \l.
		   + 
          m_K^2\,m_\pi\,\left( M_{\Lambda} + 2\,M_{N} - 3\,M_{\Sigma} \right)
              - m_\pi^3\,\left( M_{\Lambda} + 2\,M_{N} - 3\,M_{\Sigma} \right)  + 
          m_K^3\,\left( -4\,M_{\Lambda} + M_{N} + 3\,M_{\Sigma} \right)  \right) 
		  \nn \\ &&
		   + 
       2\,m_\eta\,m_K\,\left( m_K\,m_\pi^2\,
           \left( 2\,M_{\Lambda} + M_{N} - 3\,M_{\Sigma} \right)  + 
          m_K^2\,m_\pi\,\left( M_{\Lambda} + 2\,M_{N} - 3\,M_{\Sigma} \right)
              - m_\pi^3\,\left( M_{\Lambda} + 2\,M_{N} - 3\,M_{\Sigma} \right)  
		  \r. \nn \\ && \l.
			  + 
          m_K^3\,\left( -4\,M_{\Lambda} + M_{N} + 3\,M_{\Sigma} \right)  \right)  + 
       m_K^2\,\left( m_K\,m_\pi^2\,
           \left( 2\,M_{\Lambda} + M_{N} - 3\,M_{\Sigma} \right)  + 
          m_K^2\,m_\pi\,\left( M_{\Lambda} + 2\,M_{N} - 3\,M_{\Sigma} \right)
		  \r. \nn \\ && \l.
              - m_\pi^3\,\left( M_{\Lambda} + 2\,M_{N} - 3\,M_{\Sigma} \right)  + 
          m_K^3\,\left( -4\,M_{\Lambda} + M_{N} + 3\,M_{\Sigma} \right)  \right)  \Bigr] \Bigr\}
\eea	
\bea \label{eq:resp33}
&&\hspace{-12pt}\alpha^{(3)}_{\Xi^- \Lambda}=\nn \\ 
&&\frac{1}{48\,\pi \,f^2\,    {\left( m_\eta + m_K \right) }^2\,    {\left( m_K + m_\pi \right) }^2}	
\Bigl\{ -3\,{F}^2\,\Bigl[ 2\,m_\eta^3\,
        {\left( m_K + m_\pi \right) }^2 - 
       m_\eta^2\,\left( m_K^3 + 
          2\,m_K^2\,m_\pi + m_K\,m_\pi^2 - 
          2\,m_\pi^3 \right)  
		  \nn \\ &&
		  - 
       2\,m_\eta\,m_K\,
        \left( m_K^3 + 2\,m_K^2\,m_\pi + 
          m_K\,m_\pi^2 - 2\,m_\pi^3 \right)  - 
       m_K^2\,\left( m_K^3 + 
          2\,m_K^2\,m_\pi + m_K\,m_\pi^2 - 
          2\,m_\pi^3 \right)  \Bigr] \,
     \left( M_\Lambda - M_\Xi \right)  
	 \nn \\ &&
	 + 
    {D}^2\,\Bigl[ 2\,m_\eta^3\,
        {\left( m_K + m_\pi \right) }^2\,
        \left( M_\Lambda - M_\Xi \right)  + 
       m_\eta^2\,\left( m_K\,m_\pi^2\,
           \left( 2\,M_\Lambda - 3\,M_\Sigma + M_\Xi \right)  + 
          m_K^3\,\left( -4\,M_\Lambda + 3\,M_\Sigma + 
             M_\Xi \right)  
			\r. \nn \\ && \l.
			 + 
          m_K^2\,m_\pi\,
           \left( M_\Lambda - 3\,M_\Sigma + 2\,M_\Xi \right)  - 
          m_\pi^3\,\left( M_\Lambda - 3\,M_\Sigma + 
             2\,M_\Xi \right)  \right)  + 
       2\,m_\eta\,m_K\,
        \left( m_K\,m_\pi^2\,
           \left( 2\,M_\Lambda - 3\,M_\Sigma + M_\Xi \right)  
		   \r. \nn \\ && \l.
		   + 
          m_K^3\,\left( -4\,M_\Lambda + 3\,M_\Sigma + 
             M_\Xi \right)  + 
          m_K^2\,m_\pi\,
           \left( M_\Lambda - 3\,M_\Sigma + 2\,M_\Xi \right)  - 
          m_\pi^3\,\left( M_\Lambda - 3\,M_\Sigma + 
             2\,M_\Xi \right)  \right)  
			 \nn \\ && 
			 + 
       m_K^2\,\left( m_K\,m_\pi^2\,
           \left( 2\,M_\Lambda - 3\,M_\Sigma + M_\Xi \right)  + 
          m_K^3\,\left( -4\,M_\Lambda + 3\,M_\Sigma + 
             M_\Xi \right)  + 
          m_K^2\,m_\pi\,
           \left( M_\Lambda - 3\,M_\Sigma + 2\,M_\Xi \right)  
		   \r. \nn \\ && \l.
		   - 
          m_\pi^3\,\left( M_\Lambda - 3\,M_\Sigma + 
             2\,M_\Xi \right)  \right)  \Bigr]  + 
    D\,F\,\Bigl[ -4\,m_\eta^3\,
        {\left( m_K + m_\pi \right) }^2\,
        \left( M_\Lambda - M_\Xi \right)  + 
       m_\eta^2\,\left( 3\,m_\pi^3\,
           \left( 3\,M_\Lambda + M_\Sigma - 4\,M_\Xi \right) 
		  \r. \nn \\ && \l. 
		    - 
          m_K^2\,m_\pi\,
           \left( M_\Lambda + 3\,M_\Sigma - 4\,M_\Xi \right)  + 
          m_K\,m_\pi^2\,
           \left( M_\Lambda - 3\,M_\Sigma + 2\,M_\Xi \right)  + 
          m_K^3\,\left( -5\,M_\Lambda + 3\,M_\Sigma + 
             2\,M_\Xi \right)  \right) 
			\nn \\ &&			 
			  +   2\,m_\eta\,m_K\,
        \left( 3\,m_\pi^3\,
           \left( 3\,M_\Lambda + M_\Sigma - 4\,M_\Xi \right)  - 
          m_K^2\,m_\pi\,
           \left( M_\Lambda + 3\,M_\Sigma - 4\,M_\Xi \right)  + 
          m_K\,m_\pi^2\,
           \left( M_\Lambda - 3\,M_\Sigma + 2\,M_\Xi \right)  
		   \r. \nn \\ && \l.
		   + 
          m_K^3\,\left( -5\,M_\Lambda + 3\,M_\Sigma + 
             2\,M_\Xi \right)  \right)  + 
       m_K^2\,\left( 3\,m_\pi^3\,
           \left( 3\,M_\Lambda + M_\Sigma - 4\,M_\Xi \right)  - 
          m_K^2\,m_\pi\,
           \left( M_\Lambda + 3\,M_\Sigma - 4\,M_\Xi \right)  
		  \r. \nn \\ && \l. 
		   + 
          m_K\,m_\pi^2\,
           \left( M_\Lambda - 3\,M_\Sigma + 2\,M_\Xi \right)  + 
          m_K^3\,\left( -5\,M_\Lambda + 3\,M_\Sigma + 
             2\,M_\Xi \right)  \right)  \Bigr] \Bigr\}
\eea	
\bea \label{eq:resp34}
&&\hspace{-12pt}\alpha^{(3)}_{\Xi^- \Sigma^0}=\nn\\
&&\frac{1}{144\,\pi\,f^2\,{\left( m_\eta + m_K \right) }^2\,
    {\left( m_K + m_\pi \right) }^2 }
	\Bigl\{
	-3\,{F}^2\,\Bigl[ 6\,m_\eta^3\,
        {\left( m_K + m_\pi \right) }^2 + 
       m_\eta^2\,\left( m_K^3 + 
          2\,m_K^2\,m_\pi + m_K\,m_\pi^2 - 
          10\,m_\pi^3 \right)  
		  \nn \\ &&
		  + 
       2\,m_\eta\,m_K\,
        \left( m_K^3 + 2\,m_K^2\,m_\pi + 
          m_K\,m_\pi^2 - 10\,m_\pi^3 \right)  + 
       m_K^2\,\left( m_K^3 + 
          2\,m_K^2\,m_\pi + m_K\,m_\pi^2 - 
          10\,m_\pi^3 \right)  \Bigr] \,
     \left( M_\Sigma - M_\Xi \right)  
	 \nn \\ && 
	 + {D}^2\,\Bigl[  6\,m_\eta^3\,
        {\left( m_K + m_\pi \right) }^2\,
        \left( M_\Sigma - M_\Xi \right)  - 
       m_\eta^2\,\left( m_\pi^3\,
           \left( 3\,M_\Lambda + 7\,M_\Sigma - 10\,M_\Xi \right)  + 
          m_K^3\,\left( 3\,M_\Lambda - 4\,M_\Sigma + 
             M_\Xi \right)  
			 \r. \nn \\ && \l.
			 +  m_K\,m_\pi^2\,
           \left( -3\,M_\Lambda + 2\,M_\Sigma + M_\Xi \right)  + 
          m_K^2\,m_\pi\,
           \left( -3\,M_\Lambda + M_\Sigma + 2\,M_\Xi \right)  \right) 
        - 2\,m_\eta\,m_K\,
        \left( m_\pi^3\,\left( 3\,M_\Lambda + 7\,M_\Sigma - 
             10\,M_\Xi \right)  
			 \r. \nn \\ && \l.
			 + 
          m_K^3\,\left( 3\,M_\Lambda - 4\,M_\Sigma + 
             M_\Xi \right)  + 
          m_K\,m_\pi^2\,
           \left( -3\,M_\Lambda + 2\,M_\Sigma + M_\Xi \right)  + 
          m_K^2\,m_\pi\,
           \left( -3\,M_\Lambda + M_\Sigma + 2\,M_\Xi \right)  \right) 
		   \nn \\ &&
        - m_K^2\,\left( m_\pi^3\,
           \left( 3\,M_\Lambda + 7\,M_\Sigma - 10\,M_\Xi \right)  + 
          m_K^3\,\left( 3\,M_\Lambda - 4\,M_\Sigma + 
             M_\Xi \right)  + 
          m_K\,m_\pi^2\,
           \left( -3\,M_\Lambda + 2\,M_\Sigma + M_\Xi \right)  
		   \r. \nn \\ && \l.
		   + 
          m_K^2\,m_\pi\,
           \left( -3\,M_\Lambda + M_\Sigma + 2\,M_\Xi \right)  \right) 
       \Bigr]  - 3\,D\,F\,
     \Bigl[ 4\,m_\eta^3\,{\left( m_K + m_\pi \right) }^2\,
        \left( M_\Sigma - M_\Xi \right)  - 
       m_\eta^2\,\left( - m_K^2\,m_\pi\,
             \left( 3\,M_\Lambda + M_\Sigma - 4\,M_\Xi \right)
			 \r. \nn \\ && \l.
           + 3\,m_\pi^3\,\left( M_\Lambda + 3\,M_\Sigma - 
             4\,M_\Xi \right)  + 
          m_K^3\,\left( 3\,M_\Lambda - 5\,M_\Sigma + 
             2\,M_\Xi \right)  + 
          m_K\,m_\pi^2\,
           \left( -3\,M_\Lambda + M_\Sigma + 2\,M_\Xi \right)  \right) 
		   \nn \\ &&
        - 2\,m_\eta\,m_K\,
        \left( - m_K^2\,m_\pi\,
             \left( 3\,M_\Lambda + M_\Sigma - 4\,M_\Xi \right) 
           + 3\,m_\pi^3\,\left( M_\Lambda + 3\,M_\Sigma - 
             4\,M_\Xi \right)  + 
          m_K^3\,\left( 3\,M_\Lambda - 5\,M_\Sigma + 
             2\,M_\Xi \right)  
			\r. \nn \\ && \l.
			 + 
          m_K\,m_\pi^2\,
           \left( -3\,M_\Lambda + M_\Sigma + 2\,M_\Xi \right)  \right) 
        - m_K^2\,\left( - m_K^2\,m_\pi\,
             \left( 3\,M_\Lambda + M_\Sigma - 4\,M_\Xi \right) 
           + 3\,m_\pi^3\,\left( M_\Lambda + 3\,M_\Sigma - 
             4\,M_\Xi \right)  
			 \r. \nn \\ && \l.+ 
          m_K^3\,\left( 3\,M_\Lambda - 5\,M_\Sigma + 
             2\,M_\Xi \right)  + 
          m_K\,m_\pi^2\,
           \left( -3\,M_\Lambda + M_\Sigma + 2\,M_\Xi \right)  \right) 
       \Bigr] \Bigr\}
\eea
\normalsize
\section{Decuplet $\op(p^3)$ contributions}\label{sec:app2}
In this appendix we present the explicit expressions for the $\op(p^3)$ corrections
$\beta^{(3)}(\delta M_B)$ due to the octet baryon mass corrections in decuplet contributions:

\footnotesize
\bea \label{eq:resp31d} 
&&\hspace{-12pt}\beta^{(3)}_{\Sigma^- n}(\delta M_B)=\nn\\&&\frac{\cC^2}{216\,{\pi }^2\,f^2}
	\,\left.\Bigl \{ 3\,\left.\Bigl [ m_\eta^6\,\left( -8\,\delta M_N + 5\,\delta M_\Sigma \right)  - 
         18\,m_K^4\,\delta M_\Sigma\,{\Delta }^2 + 4\,m_K^2\,\left( 13\,\delta M_N - \delta M_\Sigma \right) \,{\Delta }^4 
		 \r.\r. \nn \\ && \l.\l. 
		+ 32\,\left( -\delta M_N + \delta M_\Sigma \right) \,{\Delta }^6 + 
         2\,m_\eta^4\,\left( m_K^2\,\left( 8\,\delta M_N - 11\,\delta M_\Sigma \right)  + 
            \left( 10\,\delta M_N + 11\,\delta M_\Sigma \right) \,{\Delta }^2 \right)  + 
		 \r.\r. \nn \\ && \l.\l. 
         m_\eta^2\,\left( 9\,m_K^4\,\delta M_\Sigma + 
            4\,m_K^2\,\left( -17\,\delta M_N + 11\,\delta M_\Sigma \right) \,{\Delta }^2 + 
            4\,\left( 5\,\delta M_N - 17\,\delta M_\Sigma \right) \,{\Delta }^4 \right)  \right.\Bigr] \,
       \frac{\arccos \l(\frac{\Delta }{m_\eta}\r)}{(m_\eta^2-m_K^2)^2 \sqrt{m_\eta^2-\Delta^2}}  
		 \r. \nn \\ && \l. 
		 +
      \left. 
          \left.\Bigl[ 
			- 
            12\,m_K^2\,m_\pi^2\,{\Delta }^4\,
             \left( 17\,\delta M_N\,m_\pi^2 - 5\,m_\pi^2\,\delta M_\Sigma + 16\,\delta M_N\,{\Delta }^2 - 
               16\,\delta M_\Sigma\,{\Delta }^2 \right)  
		 \r.\r.\r. \nn \\ && \l.\l. \l.			   
			   + 2\,m_K^8\,
             \left( 71\,\delta M_N\,m_\pi^2 - 80\,m_\pi^2\,\delta M_\Sigma + 91\,\delta M_N\,{\Delta }^2 + 
               98\,\delta M_\Sigma\,{\Delta }^2 \right)  + 2\,m_K^4\,{\Delta }^2\,
             \left( \delta M_N\,\left( 51\,m_\pi^4 + 416\,m_\pi^2\,{\Delta }^2 - 112\,{\Delta }^4 \right)  
		 \r.\r.\r.\r. \nn \\ && \l.\l. \l.\l.
			 - 
               2\,\delta M_\Sigma\,\left( 21\,m_\pi^4 + 64\,m_\pi^2\,{\Delta }^2 - 56\,{\Delta }^4 \right)  \right)  + 
            m_K^6\,\left( 4\,\delta M_\Sigma\,\left( 12\,m_\pi^4 + 92\,m_\pi^2\,{\Delta }^2 - 131\,{\Delta }^4 \right)
		 \r.\r.\r.\r. \nn \\ && \l.\l. \l.\l.
                   + \delta M_N\,\left( -3\,m_\pi^4 - 764\,m_\pi^2\,{\Delta }^2 + 92\,{\Delta }^4 \right)  \right)  + 
            2\,m_\eta^2\,\left( m_K^8\,\left( 41\,\delta M_N - 32\,\delta M_\Sigma \right)  - 
               6\,m_\pi^4\,\left( \delta M_N - 13\,\delta M_\Sigma \right) \,{\Delta }^4 
		 \r.\r.\r.\r. \nn \\ && \l.\l. \l.\l.
			   - 
               2\,m_K^6\,\left( 53\,\delta M_N\,m_\pi^2 - 80\,m_\pi^2\,\delta M_\Sigma + 
                  25\,\delta M_N\,{\Delta }^2 + 119\,\delta M_\Sigma\,{\Delta }^2 \right)  
		 \r.\r.\r.\r. \nn \\ && \l.\l. \l.\l.
				  - 
               m_K^4\,\left( 15\,\delta M_N\,m_\pi^4 + 48\,m_\pi^4\,\delta M_\Sigma - 
                  500\,\delta M_N\,m_\pi^2\,{\Delta }^2 + 284\,m_\pi^2\,\delta M_\Sigma\,{\Delta }^2 + 
                  302\,\delta M_N\,{\Delta }^4 - 662\,\delta M_\Sigma\,{\Delta }^4 \right)  
		 \r.\r.\r.\r. \nn \\ && \l.\l. \l.\l.
				  + 
               2\,m_K^2\,{\Delta }^2\,\left( 15\,\delta M_N\,m_\pi^4 + 21\,m_\pi^4\,\delta M_\Sigma - 
                  206\,\delta M_N\,m_\pi^2\,{\Delta }^2 - 10\,m_\pi^2\,\delta M_\Sigma\,{\Delta }^2 + 
                  160\,\delta M_N\,{\Delta }^4 - 160\,\delta M_\Sigma\,{\Delta }^4 \right)  \right)  
		 \r.\r.\r. \nn \\ && \l.\l. \l.
				  + 
            m_\eta^4\,\left( m_K^6\,\left( -47\,\delta M_N + 56\,\delta M_\Sigma \right)  + 
               2\,m_K^4\,\left( 59\,\delta M_N\,m_\pi^2 - 104\,m_\pi^2\,\delta M_\Sigma + 
                  31\,\delta M_N\,{\Delta }^2 + 68\,\delta M_\Sigma\,{\Delta }^2 \right)  
		 \r.\r.\r.\r. \nn \\ && \l.\l. \l.\l.
				  - 
               2\,{\Delta }^2\,\left( 4\,\delta M_\Sigma\,\left( 18\,m_\pi^4 + 17\,m_\pi^2\,{\Delta }^2 - 40\,{\Delta }^4 \right)
                      + \delta M_N\,\left( 9\,m_\pi^4 - 212\,m_\pi^2\,{\Delta }^2 + 160\,{\Delta }^4 \right)  \right) 
		 \r.\r.\r.\r. \nn \\ && \l.\l. \l.\l.
					   + 
               m_K^2\,\left( 8\,\delta M_\Sigma\,\left( 9\,m_\pi^4 + 61\,m_\pi^2\,{\Delta }^2 - 73\,{\Delta }^4 \right)
                      + \delta M_N\,\left( 9\,m_\pi^4 - 524\,m_\pi^2\,{\Delta }^2 + 296\,{\Delta }^4 \right)  \right)  \right) 
            \right.
		\r.\r. \nn \\ && \l.\l.\l.			
		  +m_K^{10}\,\left( -59\,\delta M_N + 32\,\delta M_\Sigma \right)  + 
            96\,m_\pi^4\,\left( \delta M_N - \delta M_\Sigma \right) \,{\Delta }^6 
		\r.	\Bigr] \,\frac{\arccos \l(\frac{\Delta }{{m_K}}\r)}{(m_\eta^2-m_K^2)^2 (m_K^2-m_\pi^2)^2\sqrt{m_K^2-\Delta^2}} 
		\r.\r. \nn \\ && \l.\l.
			\left. -2\,
             \left.\Bigl[
			 - 
               4\,\delta M_N\,\left( 7\,m_\pi^6 + 17\,m_\pi^4\,{\Delta }^2 - 73\,m_\pi^2\,{\Delta }^4 + 
                  40\,{\Delta }^6 \right)  + \delta M_\Sigma\,\left( 37\,m_\pi^6 - 58\,m_\pi^4\,{\Delta }^2 - 
                  148\,m_\pi^2\,{\Delta }^4 + 160\,{\Delta }^6 \right)  
		 \r.\r.\r.\r. \nn \\ && \l.\l. \l.\l.
				  + 
               2\,m_K^2\,\left( \delta M_\Sigma\,\left( -43\,m_\pi^4 + 158\,m_\pi^2\,{\Delta }^2 - 
                     106\,{\Delta }^4 \right)  + 2\,\delta M_N\,
                   \left( 26\,m_\pi^4 - 61\,m_\pi^2\,{\Delta }^2 + 17\,{\Delta }^4 \right)  \right)  \right.
		\r.\r.\r. \nn \\ && \l.\l. \l.		   
		 	-9\,m_K^4\,\left( 4\,\delta M_N - \delta M_\Sigma \right) \,\left( m_\pi^2 - 2\,{\Delta }^2 \right)  
				   \Bigr] \,
             \frac{\arccos \l(\frac{\Delta }{{m_\pi}}\r) }{(m_K^2-m_\pi^2)^2 \sqrt{m_\pi^2-\Delta^2}}
		 \r.\r.\r. \nn \\ && \l.\l. \l.
			 + 
            3\Delta \,
             \left.
                \left.\Bigl[ 9\,m_\eta^4\,\delta M_N - 9\,m_K^4\,\delta M_\Sigma + 
                  2\,m_K^2\,\left( 13\,\delta M_N - \delta M_\Sigma \right) \,{\Delta }^2 + 
                  16\,\left( -\delta M_N + \delta M_\Sigma \right) \,{\Delta }^4 
		 \r.\r.\r.\r.\r. \nn \\ && \l.\l. \l.\l.\l.
				  + 
                  m_\eta^2\,\left( -21\,m_K^2\,\left( \delta M_N - \delta M_\Sigma \right)  + 
                     2\,\left( \delta M_N - 13\,\delta M_\Sigma \right) \,{\Delta }^2 \right)  \right.\Bigr] \,
               \frac{ \log \l(\frac{m_\eta^2}{m_K^2}\r) }{(m_\eta^2-m_K^2)^2}
		 \r.\r.\r.\r. \nn \\ && \l.\l. \l.\l.
				- 
               \frac{2\,\Delta}{(m_\eta^2-m_K^2)(m_K^2-m_\pi^2)}
                \left. 
                   \left.\Bigl[ -\left( m_K^4\,\left( 13\,\delta M_N + 41\,\delta M_\Sigma \right)  \right)  + 
                     12\,m_\pi^2\,\left( -\delta M_N + \delta M_\Sigma \right) \,{\Delta }^2 
		 \r.\r.\r.\r.\r.\r. \nn \\ && \l.\l. \l.\l.\l.\l.					 
					 + 
                     m_\eta^2\,\left( -41\,\delta M_N\,m_\pi^2 - 13\,m_\pi^2\,\delta M_\Sigma + 
                        m_K^2\,\left( \delta M_N + 53\,\delta M_\Sigma \right)  + 40\,\delta M_N\,{\Delta }^2 - 
                        40\,\delta M_\Sigma\,{\Delta }^2 \right)  
		 \r.\r.\r.\r.\r.\r. \nn \\ && \l.\l. \l.\l.\l.\l.					 
						+ 
                     m_K^2\,\left( 53\,\delta M_N\,m_\pi^2 + m_\pi^2\,\delta M_\Sigma - 
                        28\,\delta M_N\,{\Delta }^2 + 28\,\delta M_\Sigma\,{\Delta }^2 \right)  \right.\Bigr]
		 \r.\r.\r.\r.\r. \nn \\ && \l.\l. \l.\l.\l. 
                  -2\Delta \,
                   \left.\Bigl[ -9\,\delta M_N\,m_\pi^4 + 36\,m_\pi^4\,\delta M_\Sigma + 
                     9\,m_K^4\,\left( -4\,\delta M_N + \delta M_\Sigma \right)  - 106\,\delta M_N\,m_\pi^2\,{\Delta }^2 + 
                     34\,m_\pi^2\,\delta M_\Sigma\,{\Delta }^2 
		 \r.\r.\r.\r.\r.\r. \nn \\ && \l.\l.\l. \l.\l.\l.					
					 + 80\,\delta M_N\,{\Delta }^4 - 80\,\delta M_\Sigma\,{\Delta }^4 + 
                     m_K^2\,\left( 105\,\delta M_N\,m_\pi^2 - 105\,m_\pi^2\,\delta M_\Sigma - 
                        34\,\delta M_N\,{\Delta }^2 + 106\,\delta M_\Sigma\,{\Delta }^2 \right)  \right.\Bigr] \,
                   \frac{\log \l(\frac{m_K^2}{m_\pi^2}\r)}{(m_K^2-m_\pi^2)^2} \Bigr\} \,,\right.  \right.  \right.  \right. \right.
\eea
\bea \label{eq:resp32d} 
&&\hspace{-12pt}\beta^{(3)}_{\Lambda p}(\delta M_B)=\nn\\&&-\frac{\cC^2}{72\,{\pi }^2\,f^2\,
    {\left( m_K^2 - m_\pi^2 \right) }^2 }\,
		 \nn \\ && 
	\left.\Bigl\{ 
         \left.\Bigl[ m_K^6\,\left( -8\,\delta M_\Lambda + 5\,\delta M_N \right)  - 18\,\delta M_N\,m_\pi^4\,{\Delta }^2 + 
           52\,\delta M_\Lambda\,m_\pi^2\,{\Delta }^4 - 4\,\delta M_N\,m_\pi^2\,{\Delta }^4 - 32\,\delta M_\Lambda\,{\Delta }^6 + 
           32\,\delta M_N\,{\Delta }^6 
		   		\r. \r. \nn \\ && \l.\l.
				+ 2\,m_K^4\,
            \left( 11\,\delta M_N\,\left( -m_\pi^2 + {\Delta }^2 \right)  + 
              2\,\delta M_\Lambda\,\left( 4\,m_\pi^2 + 5\,{\Delta }^2 \right)  \right) 
		 \r.\r. \nn \\ &&\l. \l.
			   + 
           m_K^2\,\left( -68\,\delta M_\Lambda\,m_\pi^2\,{\Delta }^2 + 20\,\delta M_\Lambda\,{\Delta }^4 + 
              \delta M_N\,\left( 9\,m_\pi^4 + 44\,m_\pi^2\,{\Delta }^2 - 68\,{\Delta }^4 \right)  \right)  \right.\Bigr] \,
         \frac{\arccos \l(\frac{\Delta }{{m_K}}\r)}{{\sqrt{m_K^2 - {\Delta }^2}}}
		 \r. \nn \\ && \l. 
        \left. \left. + \Bigl[ 9\,m_K^4\,\delta M_\Lambda\,
               \left( m_\pi^2 - 2\,{\Delta }^2 \right)  + 
              4\,\delta M_N\,\left( -2\,m_\pi^6 + 5\,m_\pi^4\,{\Delta }^2 + 5\,m_\pi^2\,{\Delta }^4 - 
                 8\,{\Delta }^6 \right)  
		 \r.\r.\r. \nn \\ &&\l. \l.\l.				 
				 + \delta M_\Lambda\,\left( 5\,m_\pi^6 + 22\,m_\pi^4\,{\Delta }^2 - 
                 68\,m_\pi^2\,{\Delta }^4 + 32\,{\Delta }^6 \right)  
		 \r.\r.\r. \nn \\ && \l.\l.\l.
 				 + 
              m_K^2\,\left( \delta M_\Lambda\,\left( -22\,m_\pi^4 + 44\,m_\pi^2\,{\Delta }^2 - 4\,{\Delta }^4 \right)
                     + 4\,\delta M_N\,\left( 4\,m_\pi^4 - 17\,m_\pi^2\,{\Delta }^2 + 13\,{\Delta }^4 \right)  \right)  \right.\Bigr] 
             \,\frac{\arccos \l(\frac{\Delta }{{m_\pi}}\r)}{{\sqrt{m_\pi^2 - {\Delta }^2}}}
		\r.\r. \nn \\ && \l.\l.
			  + 
           2\Delta (m_K^2-m_\pi^2) \,
            \left. 
               \left.\Bigl[ -\left( m_K^2\,\left( \delta M_\Lambda + 5\,\delta M_N \right)  \right)  + 5\,\delta M_\Lambda\,m_\pi^2 + 
                 \delta M_N\,m_\pi^2 - 4\,\delta M_\Lambda\,{\Delta }^2 + 4\,\delta M_N\,{\Delta }^2 \right.\Bigr]  \
		 \r.\r.\r. \nn \\ && \l.\l.\l.				 
				 + 
              \left.\Delta\Bigl[ 9\,m_K^4\,\delta M_\Lambda - 9\,\delta M_N\,m_\pi^4 + 
                 26\,\delta M_\Lambda\,m_\pi^2\,{\Delta }^2 - 2\,\delta M_N\,m_\pi^2\,{\Delta }^2 - 
                 16\,\delta M_\Lambda\,{\Delta }^4 + 16\,\delta M_N\,{\Delta }^4 
		 \r.\r.\r.\r. \nn \\ && \l.\l.\l.\l.				 
				 + 
                 m_K^2\,\left( -21\,\delta M_\Lambda\,m_\pi^2 + 21\,\delta M_N\,m_\pi^2 + 
                    2\,\delta M_\Lambda\,{\Delta }^2 - 26\,\delta M_N\,{\Delta }^2 \right)  \right.\Bigr] \,
               \log \l(\frac{m_K^2}{m_\pi^2}\r) \right.  \right.  \right.\Bigr\}\,,
\eea
\bea \label{eq:resp33d} 	
&&\hspace{-12pt}\beta^{(3)}_{\Xi^- \Lambda}(\delta M_B)=\nn\\&&\frac{\cC^2}{72\,{\pi }^2\,f^2}
		\,\left.\Bigl\{ \left.\Bigl[ m_\eta^6\,\left( -8\,\delta M_\Lambda + 5\,\delta M_\Xi \right)  - 
         18\,m_K^4\,\delta M_\Xi\,{\Delta }^2 + 4\,m_K^2\,\left( 13\,\delta M_\Lambda - \delta M_\Xi \right) \,{\Delta }^4 
		\r.\r. \nn \\ && \l.\l.
		 + 
         32\,\left( -\delta M_\Lambda + \delta M_\Xi \right) \,{\Delta }^6 + 
         2\,m_\eta^4\,\left( m_K^2\,\left( 8\,\delta M_\Lambda - 11\,\delta M_\Xi \right)  + 
            \left( 10\,\delta M_\Lambda + 11\,\delta M_\Xi \right) \,{\Delta }^2 \right)  
		\r.\r. \nn \\ && \l.\l.			
			+ 
         m_\eta^2\,\left( 9\,m_K^4\,\delta M_\Xi + 
            4\,m_K^2\,\left( -17\,\delta M_\Lambda + 11\,\delta M_\Xi \right) \,{\Delta }^2 + 
            4\,\left( 5\,\delta M_\Lambda - 17\,\delta M_\Xi \right) \,{\Delta }^4 \right)  \right.\Bigr] \,
       \frac{\arccos \l(\frac{\Delta }{{m_\eta}}\r) }{\sqrt{m_\eta^2 - {\Delta }^2}{\left( m_\eta^2 - m_K^2 \right) }^2}
		\r. \nn \\ && \l. 
      \left. 
          \left. + \Bigl[ m_K^{10}\,\left( -8\,\delta M_\Lambda + 5\,\delta M_\Xi \right)  + 
            32\,m_\pi^4\,\left( \delta M_\Lambda - \delta M_\Xi \right) \,{\Delta }^6
		 \r.\r.\r. \nn \\ && \l.\l.\l.				 
			 + 
            4\,m_K^2\,m_\pi^2\,{\Delta }^4\,
             \left( -17\,\delta M_\Lambda\,m_\pi^2 + 5\,m_\pi^2\,\delta M_\Xi - 16\,\delta M_\Lambda\,{\Delta }^2 + 
               16\,\delta M_\Xi\,{\Delta }^2 \right)  
		 \r.\r.\r. \nn \\ && \l.\l.\l.				 
			   + m_K^8\,
             \left( 22\,\delta M_\Xi\,\left( -m_\pi^2 + {\Delta }^2 \right)  + 
               4\,\delta M_\Lambda\,\left( 7\,m_\pi^2 + 5\,{\Delta }^2 \right)  \right)  + 
            m_K^6\,\left( \delta M_\Xi\,\left( m_\pi^4 + 72\,m_\pi^2\,{\Delta }^2 - 68\,{\Delta }^4 \right)  
		 \r.\r.\r.\r. \nn \\ && \l.\l.\l.\l.				 
			- 
               4\,\delta M_\Lambda\,\left( m_\pi^4 + 39\,m_\pi^2\,{\Delta }^2 - 5\,{\Delta }^4 \right)  \right)  + 
            2\,m_K^4\,{\Delta }^2\,\left( 4\,\delta M_\Lambda\,
                \left( 5\,m_\pi^4 + 24\,m_\pi^2\,{\Delta }^2 - 4\,{\Delta }^4 \right)  
		\r.\r.\r.\r. \nn \\ && \l.\l.\l.\l.
				+ 
               \delta M_\Xi\,\left( m_\pi^4 - 48\,m_\pi^2\,{\Delta }^2 + 16\,{\Delta }^4 \right)  \right)  
		 \r.\r.\r. \nn \\ && \l.\l.\l.				 
			   + 
            m_\eta^4\,\left( m_K^6\,\left( -4\,\delta M_\Lambda + 13\,\delta M_\Xi \right)  - 
               18\,m_\pi^4\,\delta M_\Xi\,{\Delta }^2 + 
               56\,m_\pi^2\,\left( \delta M_\Lambda - \delta M_\Xi \right) \,{\Delta }^4 + 
               64\,\left( -\delta M_\Lambda + \delta M_\Xi \right) \,{\Delta }^6 
		 \r.\r.\r.\r. \nn \\ && \l.\l.\l.\l.				 
			   + 
               m_K^2\,\left( 9\,m_\pi^4\,\delta M_\Xi + 
                  4\,m_\pi^2\,\left( -19\,\delta M_\Lambda + 28\,\delta M_\Xi \right) \,{\Delta }^2 + 
                  88\,\left( \delta M_\Lambda - \delta M_\Xi \right) \,{\Delta }^4 \right)  
		 \r.\r.\r.\r. \nn \\ && \l.\l.\l.\l.				 
				  + 
               2\,m_K^4\,\left( 10\,\delta M_\Lambda\,\left( m_\pi^2 - {\Delta }^2 \right)  + 
                  \delta M_\Xi\,\left( -19\,m_\pi^2 + {\Delta }^2 \right)  \right)  \right)  + 
            2\,m_\eta^2\,\left( m_K^8\,\left( 2\,\delta M_\Lambda - 5\,\delta M_\Xi \right)  
		 \r.\r.\r.\r. \nn \\ && \l.\l.\l.\l.				 
			- 
               2\,m_\pi^4\,\left( \delta M_\Lambda - 13\,\delta M_\Xi \right) \,{\Delta }^4 + 
               m_K^6\,\left( -16\,\delta M_\Lambda\,m_\pi^2 + 22\,m_\pi^2\,\delta M_\Xi + 
                  24\,\delta M_\Lambda\,{\Delta }^2 - 36\,\delta M_\Xi\,{\Delta }^2 \right) 
		 \r.\r.\r.\r. \nn \\ && \l.\l.\l.\l.				 
				   + 
               4\,m_K^2\,{\Delta }^2\,\left( \delta M_\Xi\,
                   \left( -4\,m_\pi^4 + m_\pi^2\,{\Delta }^2 - 16\,{\Delta }^4 \right)  + 
                  \delta M_\Lambda\,\left( m_\pi^4 - 13\,m_\pi^2\,{\Delta }^2 + 16\,{\Delta }^4 \right)  \right) 
		 \r.\r.\r.\r. \nn \\ && \l.\l.\l.\l.				 
				   - 
               m_K^4\,\left( \delta M_\Xi\,\left( m_\pi^4 + 44\,m_\pi^2\,{\Delta }^2 - 114\,{\Delta }^4 \right)  + 
                  2\,\delta M_\Lambda\,\left( m_\pi^4 - 34\,m_\pi^2\,{\Delta }^2 + 45\,{\Delta }^4 \right)  \right)  \right)  \right.\Bigr]\times
		\r.\r. \nn \\ && \l.\l.				 		  
            \times\,\frac{\arccos \l(\frac{\Delta }{{m_K}}\r)}{{\sqrt{m_K^2 - {\Delta }^2}}{\left( m_\eta^2 - m_K^2 \right) }^2\,{\left( m_K^2 - m_\pi^2 \right) }^2}
		\r.\r. \nn \\ && \l.\l.				 
			 + 
         \left. 
             \left( \delta M_\Lambda - \delta M_\Xi \right) \,
             \left.\Bigl[ 13\,m_\pi^6 + 2\,m_\pi^4\,{\Delta }^2 - 88\,m_\pi^2\,{\Delta }^4 + 64\,{\Delta }^6 + 
               9\,m_K^4\,\left( m_\pi^2 - 2\,{\Delta }^2 \right)  			   
		 \r.\r.\r.\r. \nn \\ && \l.\l.\l.\l.				 
			   - 
               2\,m_K^2\,\left( 19\,m_\pi^4 - 56\,m_\pi^2\,{\Delta }^2 + 28\,{\Delta }^4 \right)  \right.\Bigr] \,
             \frac{\arccos \l(\frac{\Delta }{m_\pi}\r)}{{\sqrt{m_\pi^2 - {\Delta }^2}}{\left( m_K^2 - m_\pi^2 \right) }^2}
         \r.\r.\r. \nn \\ && \l.\l.\l.			 
			  + 
            \Delta
             \left. 
                \left.\Bigl[ 9\,m_\eta^4\,\delta M_\Lambda - 9\,m_K^4\,\delta M_\Xi + 
                  2\,m_K^2\,\left( 13\,\delta M_\Lambda - \delta M_\Xi \right) \,{\Delta }^2 
		 \r.\r.\r.\r.\r. \nn \\ && \l.\l.\l.\l.\l.				 
				  + 
                  16\,\left( -\delta M_\Lambda + \delta M_\Xi \right) \,{\Delta }^4 + 
                  m_\eta^2\,\left( -21\,m_K^2\,\left( \delta M_\Lambda - \delta M_\Xi \right)  + 
                     2\,\left( \delta M_\Lambda - 13\,\delta M_\Xi \right) \,{\Delta }^2 \right)  \right.\Bigr] \,
                \frac{\log \l(\frac{m_\eta^2}{m_K^2}\r) }{{\left( m_\eta^2 - m_K^2 \right) }^2}
		 \r.\r.\r.\r. \nn \\ && \l.\l.\l.\l.				 
				+ 
                \,
                \left. \frac{2\,\Delta}{{\left( m_\eta^2 - m_K^2 \right) }\,{\left( m_K^2 - m_\pi^2 \right) }}
                   \left.\Bigl[ m_K^4\,\left( \delta M_\Lambda + 5\,\delta M_\Xi \right)  + 
                     4\,m_\pi^2\,\left( \delta M_\Lambda - \delta M_\Xi \right) \,{\Delta }^2 
		 \r.\r.\r.\r.\r.\r. \nn \\ && \l.\l.\l.\l.\l.\l.				 
					 + 
                     m_K^2\,\left( -9\,\delta M_\Lambda\,m_\pi^2 + 3\,m_\pi^2\,\delta M_\Xi + 
                        4\,\delta M_\Lambda\,{\Delta }^2 - 4\,\delta M_\Xi\,{\Delta }^2 \right)  
		 \r.\r.\r.\r.\r.\r. \nn \\ && \l.\l.\l.\l.\l.\l.				 
						+ 
                     m_\eta^2\,\left( 5\,\delta M_\Lambda\,m_\pi^2 + 
                        3\,m_K^2\,\left( \delta M_\Lambda - 3\,\delta M_\Xi \right)  + m_\pi^2\,\delta M_\Xi - 
                        8\,\delta M_\Lambda\,{\Delta }^2 + 8\,\delta M_\Xi\,{\Delta }^2 \right)  \right.\Bigr] 
		 \r.\r.\r.\r.\r. \nn \\ && \l.\l.\l.\l.\l.
						 + 
                  \Delta\left( \delta M_\Lambda - \delta M_\Xi \right) \,
                   \left.\Bigl[ 9\,m_K^4 + 9\,m_\pi^4 + 28\,m_\pi^2\,{\Delta }^2 - 32\,{\Delta }^4 + 
                     m_K^2\,\left( -42\,m_\pi^2 + 28\,{\Delta }^2 \right)  \right.\Bigr] \,
                   \frac{\log \l(\frac{m_K^2}{m_\pi^2}\r)}{{\left( m_K^2 - m_\pi^2 \right) }^2} \right.  \right.  \right.  \right.  \right.\Bigr\} \,,
\eea
\bea\label{eq:resp34d} 
&&\hspace{-12pt}\beta^{(3)}_{\Xi^- \Sigma^0}(\delta M_B)=\nn\\&&-\frac{ \cC^2}{216\,{\pi }^2\,f^2}	
	\left.\Bigl\{ 9\,
         \left( \delta M_\Sigma + \delta M_\Xi \right) \,
         \left.\Bigl[ -m_\eta^4 - 3\,m_\eta^2\,m_K^2 + 14\,m_\eta^2\,{\Delta }^2 + 
           6\,m_K^2\,{\Delta }^2 - 16\,{\Delta }^4 \right.\Bigr] \,
		   \frac{\arccos \l(\frac{\Delta }{m_\eta}\r)}{\left( m_\eta^2 - m_K^2 \right)\sqrt{m_\eta^2 - {\Delta }^2}}
		    \r. \nn \\ && \l.
        \left. -\left. 
              \left.\Bigl[ -\left( m_K^8\,\left( 20\,\delta M_\Sigma + 7\,\delta M_\Xi \right)  \right)  - 
                144\,m_\pi^4\,\left( \delta M_\Sigma + \delta M_\Xi \right) \,{\Delta }^4
		 \r.\r.\r.\r. \nn \\ && \l.\l.\l.\l.				 
				 + 
                2\,m_K^6\,\left( 14\,\delta M_\Sigma\,m_\pi^2 - 5\,m_\pi^2\,\delta M_\Xi + 
                   94\,\delta M_\Sigma\,{\Delta }^2 + 95\,\delta M_\Xi\,{\Delta }^2 \right)  
		 \r.\r.\r.\r. \nn \\ && \l.\l.\l.\l.				 
				   + 
                m_K^4\,\left( 9\,m_\pi^4\,\delta M_\Xi - 
                   8\,m_\pi^2\,\left( 43\,\delta M_\Sigma + 29\,\delta M_\Xi \right) \,{\Delta }^2 - 
                   4\,\left( 43\,\delta M_\Sigma + 65\,\delta M_\Xi \right) \,{\Delta }^4 \right) 
		 \r.\r.\r.\r. \nn \\ && \l.\l.\l.\l.				 
				    + 
                2\,m_K^2\,{\Delta }^2\,\left( 2\,\delta M_\Sigma\,
                    \left( 27\,m_\pi^4 + 97\,m_\pi^2\,{\Delta }^2 - 8\,{\Delta }^4 \right)  + 
                   \delta M_\Xi\,\left( 45\,m_\pi^4 + 166\,m_\pi^2\,{\Delta }^2 + 16\,{\Delta }^4 \right)  \right)  
		 \r.\r.\r.\r. \nn \\ && \l.\l.\l.\l.				 
				   - 
                m_\eta^2\,\left( m_K^6\,\left( 16\,\delta M_\Sigma + 29\,\delta M_\Xi \right)  - 
                   2\,{\Delta }^2\,\left( 9\,m_\pi^2 - 8\,{\Delta }^2 \right) \,
                    \left( 4\,\delta M_\Sigma\,m_\pi^2 + 5\,m_\pi^2\,\delta M_\Xi - 2\,\delta M_\Sigma\,{\Delta }^2 + 
                      2\,\delta M_\Xi\,{\Delta }^2 \right)  
		 \r.\r.\r.\r.\r. \nn \\ && \l.\l.\l.\l.\l.
					  + 
                   m_K^4\,\left( -44\,\delta M_\Sigma\,m_\pi^2 - 82\,m_\pi^2\,\delta M_\Xi + 
                      8\,\delta M_\Sigma\,{\Delta }^2 + 10\,\delta M_\Xi\,{\Delta }^2 \right)  + 
                   m_K^2\,\left( \delta M_\Xi\,
                       \left( 45\,m_\pi^4 + 128\,m_\pi^2\,{\Delta }^2 - 116\,{\Delta }^4 \right)  
		 \r.\r.\r.\r.\r.\r. \nn \\ && \l.\l.\l.\l.\l.\l.
					   + 
                      4\,\delta M_\Sigma\,\left( 9\,m_\pi^4 + 4\,m_\pi^2\,{\Delta }^2 - 7\,{\Delta }^4 \right)  \right)  \right) 
                \right.\Bigr] \,\frac{\arccos \l(\frac{\Delta }{m_K}\r)}{\sqrt{m_K^2 - {\Delta }^2}\left( m_\eta^2 - m_K^2 \right) \,{\left( m_K^2 - m_\pi^2 \right) }^2}
				 \right.  
				\r.\r. \nn \\ && \l.\l.
				- 
           \left. 
               \left.\Bigl[ -9\,m_K^4\,\left( 2\,\delta M_\Sigma + \delta M_\Xi \right) \,\left( m_\pi^2 - 2\,{\Delta }^2 \right)  - 
                 2\,\delta M_\Sigma\,\left( m_\pi^6 + 32\,m_\pi^4\,{\Delta }^2 - 58\,m_\pi^2\,{\Delta }^4 + 
                    16\,{\Delta }^6 \right) 
		 \r.\r.\r.\r. \nn \\ && \l.\l.\l.\l.				 
					 + 
					 \delta M_\Xi\,
                  \left( 11\,m_\pi^6 - 62\,m_\pi^4\,{\Delta }^2 + 28\,m_\pi^2\,{\Delta }^4 + 32\,{\Delta }^6 \right)  - 
                 2\,m_K^2\,\left( \delta M_\Sigma\,
                     \left( -14\,m_\pi^4 + 10\,m_\pi^2\,{\Delta }^2 + 22\,{\Delta }^4 \right)  
		 \r.\r.\r.\r.\r. \nn \\ && \l.\l.\l.\l.\l.				 
					 + 
                    \delta M_\Xi\,\left( 5\,m_\pi^4 - 46\,m_\pi^2\,{\Delta }^2 + 50\,{\Delta }^4 \right)  \right)  \right.\Bigr] \,
               \frac{\arccos \l(\frac{\Delta }{m_\pi}\r)}{{\sqrt{m_\pi^2 - {\Delta }^2}}{\left( m_K^2 - m_\pi^2 \right) }^2} 
			   \r.\r.\r. \nn \\ && \l.\l.\l.
			   + 
              9\Delta 
               \left. \left( \delta M_\Sigma + \delta M_\Xi \right) \,
                  \left.\Bigl[ 3\,m_\eta^2 + 3\,m_K^2 - 8\,{\Delta }^2 \right.\Bigr] \,
				  \frac{\log \l(\frac{m_\eta^2}{m_K^2}\r) }{ \left( m_\eta^2 - m_K^2 \right)}
		 \r.\r.\r.\r. \nn \\ && \l.\l.\l.\l.				 
				  - \frac{2 \Delta}{\left( m_K^2 - m_\pi^2 \right)}
                  \left. 
                     \left.\Bigl[ -29\,\delta M_\Sigma\,m_\pi^2 - 25\,m_\pi^2\,\delta M_\Xi + 
                       m_K^2\,\left( 25\,\delta M_\Sigma + 29\,\delta M_\Xi \right)  + 4\,\delta M_\Sigma\,{\Delta }^2 - 
                       4\,\delta M_\Xi\,{\Delta }^2 \right.\Bigr]
		 \r.\r.\r.\r.\r. \nn \\ && \l.\l.\l.\l.\l.				 
					    -\Delta
                    \left.\Bigl[ -9\,m_K^4\,\left( 2\,\delta M_\Sigma + \delta M_\Xi \right)  + 
                       m_K^2\,\left( 21\,\delta M_\Sigma\,m_\pi^2 - 21\,m_\pi^2\,\delta M_\Xi + 
                          22\,\delta M_\Sigma\,{\Delta }^2 + 50\,\delta M_\Xi\,{\Delta }^2 \right)  
		 \r. \r.\r.\r.\r.\r. \nn \\ && \l.\l.\l.\l.\l.\l.				 
						  + 
                       2\,\delta M_\Xi\,\left( 9\,m_\pi^4 - 11\,m_\pi^2\,{\Delta }^2 - 8\,{\Delta }^4 \right)  + 
                       \delta M_\Sigma\,\left( 9\,m_\pi^4 - 50\,m_\pi^2\,{\Delta }^2 + 16\,{\Delta }^4 \right)  \right.\Bigr] \,
                     \frac{\log \l(\frac{m_K^2}{m_\pi^2}\r)}{{\left( m_K^2 - m_\pi^2 \right) }^2} \right.  \right.  \right.  \right.  \right.\Bigr\}\,.
\eea

\end{document}